\documentclass[aps,pre,onecolumn,superscriptaddress,floatfix]{revtex4}
\usepackage{graphicx,color,epsfig}
\usepackage{color}
\usepackage{epstopdf}
\usepackage[utf8]{inputenc}

\begin{document}
\title{Propagation of Systemic Risk in Interbank Networks}

\author{Vanessa Hoffmann de Quadros}

\affiliation{Instituto de Física, Universidade Federal do Rio Grande
  do Sul, Caixa Postal 15051, 90501-970 Porto Alegre RS, Brazil}

\author{Juan Carlos González-Avella}

\affiliation{Instituto de Física, Universidade Federal do Rio Grande
  do Sul, Caixa Postal 15051, 90501-970 Porto Alegre RS, Brazil}

\author{José Roberto Iglesias}

\affiliation{Instituto de Física, Universidade Federal do Rio Grande
  do Sul, Caixa Postal 15051, 90501-970 Porto Alegre RS, Brazil}

\affiliation{Programa de Mestrado em Economia, Universidade do Vale do Rio dos Sinos,
Av. Unisinos, 950, 93022-000 São Leopoldo RS, Brazil}
\date{\today}

\begin{abstract}
\vspace{1em}
\noindent
\textbf{Abstract}
One of the most striking characteristics of modern financial systems is its complex interdependence, standing out the network of bilateral exposures in
interbank market, through which institutions with surplus liquidity can lend to those with liquidity shortage. While the interbank market is responsible
for efficient liquidity allocation, it also introduces the possibility for systemic risk via financial contagion. Insolvency of one bank can propagate
through the network leading to insolvency of other banks. Moreover, empirical studies reveal that some interbank networks have features of scale-free
networks, which means that the distribution of connections among banks follows a power law. This work explores the characteristics of financial contagion in networks whose links distributions approaches a power law, using a model that defines
banks balance sheets from information of network connectivity. By varying the parameters for the creation of the network, several interbank networks are built, in which the concentrations of debts and credits are
obtained from links distributions during the creation networks process. Three main types of interbank network are analyzed for their resilience to contagion:
i)  concentration of debts is greater than concentration of credits, ii) concentration of credits is greater than concentration of debts and
iii) concentrations of debts and credits are similar. We also tested the effect of a variation in connectivity in conjunction with variation in
concentration of links. The results suggest that more connected networks with high concentration of credits (featuring nodes that are large creditors of the system) present greater
resilience to contagion when compared with the others networks analyzed. Evaluating some topological indices of systemic risk suggested by the literature we have verified the ability of these indices to explain the impact on
the system caused by the failure of a node. There is a clear positive correlation between the topological indices and the magnitude of losses in the
case of networks with high concentration of debts. This correlation is smaller for more resilient networks.
\vspace{1em}

\noindent
\textbf{Resumo}
\noindent
Uma das características mais marcantes dos sistemas financeiros modernos é sua complexa interdependência, destacando-se a rede de
exposições bilaterais no mercado interbancário, através do qual as instituições com liquidez excedente podem emprestar para aqueles com 
falta de liquidez. Enquanto o mercado interbancário é responsável pela alocação eficiente de liquidez, também introduz a possibilidade de
risco sistêmico através de contágio financeiro. A insolvência de um banco pode se propagar através da rede, levando à insolvência outros bancos.
Além disso, os estudos empíricos revelam que algumas redes interbancárias têm características de redes livres de escala, que significa que a distribuição 
das conexões entre os bancos segue uma lei de potências. Este trabalho explora as características do contágio financeiro em redes cuja distribuição de 
conexões se aproxima de uma lei de potência, utilizando um modelo que define o balance financeiro dos bancos usando as informações de conectividade de rede.
Variando os parâmetros para a criação da rede, várias redes interbancárias são construídas, em que as concentrações de débitos e créditos 
são obtidas a partir de distribuições de ligações durante o processo de construção das redes. Três tipos principais de rede interbancária são 
analisados segundo sua resiliência ao contágio: i) concentração de dívidas maior que a concentração de créditos, ii) concentração de créditos maior 
do que a concentração de dívidas e iii) concentrações de dívidas e créditos semelhantes. Testamos também o efeito de uma variação na conectividade
em conjunto com a variação na concentração de links.
Os resultados sugerem que redes mais conectadas e com alta concentração de créditos (apresentando nós que são os grandes credores do sistema) 
apresentam maior resiliência ao contágio quando comparado com as outras redes analisadas. Avaliando alguns índices topológicos de risco 
sistémico sugeridos pela literatura podemos verificar a capacidade destes índices para explicar o impacto no sistema causado pela falha 
de um nó. Há uma correlação positiva entre os índices topológicos e a magnitude das perdas no caso de redes com alta concentração de dívidas. 
Esta correlação é menor para redes mais resistentes.


\noindent
\textbf{Palavras chave:} Exposição interbancária, leis de potência, risco sistémico, redes complexas, crisis financeiras.


\noindent
\textbf{\footnotesize Classificação ANPEC:} Área 8

\end{abstract}

\maketitle

\section{Introduction}\label{intro}

The financial crisis started in 2007 highlighted, once again, the high degree of interdependence of financial systems. A combination of excessive borrowing,
risky investments, lack of transparency and high interdependence led the financial system to the worst financial meltdown since the Great Depression (FCIC \cite{FCIC}). An increasing interest in financial contagion, motivated by the crisis, gave rise to several works in this field in
the last years, although the subject is not new.

The interdependence of financial systems exhibits multiple channels. Financial institutions are connected through mutual exposures created in the interbank
market, through which institutions with surplus liquidity can lend to those with liquidity shortage. While the interbank market is responsible for efficient
liquidity allocation, it also introduces the possibility for credit risk via financial contagion. Insolvency of one bank can propagate through its links leading
to insolvency of other banks. Equally important, financial institutions are indirectly connected by holding similar assets exposures and by sharing the same
mass of depositors.

With respect to the direct connections of mutual exposures, the structure of interdependence can be easily captured by using a network representation, in
which the nodes of the network represent financial institutions, while the links represent exposures between nodes (Allen and Babus \cite{Babus2009}).
Mutual exposures networks are directed networks in which {\it in} links stand for credits while {\it out} links represent debts. The link direction
indicates the cash flow at the time of debt payment (from debtor to creditor) and also indicates the direction of impact or financial loss in case of
default of borrowers.

Theoretical studies (Allen and Gale \cite{Allen.Gale2000FinancialContagion}, Freixas et. al. \cite{Freixas2000}) have shown that the possibility for contagion
via mutual exposures depends on the precise structure of the interbank market. Allen and Gale \cite{Allen.Gale2000FinancialContagion}
assess the contagion generated from an increased demand for liquidity (liquidity shock) of a system consisting of 4 identical regions. They demonstrate that
if the system formed by these 4 regions is a complete network, in which all nodes have connections to each other, then contagion effect is minimized. The
authors argue that in complete networks the impact of a liquidity crisis in one region may be distributed among all others and thus attenuated. Similar
results are found by Freixas et. al. \cite{Freixas2000} in a model with money-centre banks.

In recent studies, different models of complex network have been used to generate artificial interbank networks, in order to identify if a given network is
more or less prone to contagion.

Nier et al. \cite{Nier.Yang.ea2007Networkmodelsand} simulate contagion from the initial failure of a bank in an Erdös-Rényi random network, finding a
negative nonlinear relationship between
contagion and the level of capitalization of banks. The relationship between contagion and connectivity is also nonlinear. An increase in the number of
interbank exposures initially has no effect on contagion, since the losses are absorbed by the capital of each affected node. However, as the number of
connections is raised, contagion increases until the point where further increase in connectivity cause contagion to decline. The non-monotonic
relationship between connectivity and contagion found by the authors reflects the action of two effects: on the one hand, the addition of new links adds
new channels through which contagion can occur. On the other hand, increasing links also represent the distribution of losses among a larger number of
nodes, diluting the impact of the failure and mitigating the effects of the crisis.

Studying a specific case of power-law network, Cont and Moussa \cite{Rama2010b} find similar results as Nier et al. \cite{Nier.Yang.ea2007Networkmodelsand}
for the relation between connectivity, level of capitalization and contagion.

Battiston et al. \cite{Battiston2009} simulate contagion on a regular network and also find a nonlinear relationship between connectivity and contagion, but
with opposite effect: initially the increase in the number of connections decreases the network contagion, while later additions make contagion to increase.

The differences in the results indicate that the possibility and extent of contagion depends considerably on the structure of each network and specific
assumptions of each model.

Empirical studies reveal that some interbank networks have features of scale-free networks: this means that the distribution of connections among banks
follows a power law, $p(k) \sim k^{-X}$ (Boss et al. \cite{Boss2004}, Cont et al. \cite{Rama2010}, Soramäki et al. \cite{RePEc:fip:fednsr:243}, Inaoka et
al. \cite{Inaoka2004}). In general terms, some of the most outstanding features reported in the literature are:

\begin{itemize}
 \item Networks have low density of links, that is, they are far from complete;
 \item Asymmetrical in-degree and out-degree distribution;
 \item Power law distributions for in and out-degree whose exponent varies around 2 and 3.
\end{itemize}

It is also noteworthy the characteristic reported by Cont et al. \cite{Rama2010} in the study of the Brazilian network: there is a positive association between
the size of exposures (assets) and the number of debtors (in-degree) of an institution, and a positive association between the size of liabilities and the
number of creditors (out-degree) of an institution. More (less) connected financial institutions have larger (smaller) exposures (links weights).

The aim of this paper is to identify, through simulations of networks whose distributions approach power laws, how scale free networks behave with regard to
financial contagion via mutual exposures and what characteristics make a given network more or less prone to propagate crises. Our particular interest is
to evaluate the role of the exponents that characterize the scale free network, since these exponents determine the concentration of debts (out-degree) and
credits (in-degree) in the financial network. We construct networks whose connectivity distribution approaches a
power law using the algorithm introduced by Bollobás et al. \cite{conf/soda/BollobasBCR03}, as extension of the method for network construction used by
Barabasi and Albert \cite{barabasi99}. Another important ingredient that we include in our analysis is to introduce a simplified model that determines the
banks balance sheets from information of network connectivity; we simulate the patrimonial structure of the financial system by dividing each balance sheet
in bank assets and liabilities (relationships with other financial agents) and nonbank assets and liabilities (relations with non-financial agents).

By varying the parameters for the creation of the network, several interbank networks are built, in which concentrations of debts and credits are
obtained from links distributions during the networks creation process. Three main types of interbank network are analyzed for their resilience to
contagion:
i) networks in which concentration of debts is greater than concentration of credits, ii) networks in which concentration of credits is greater than
concentration of debts and iii) networks in which concentrations of debts and credits are similar. We also tested the effect of a variation in connectivity
in conjunction with variation in concentration of links, as well as the effect of network size and the level of bank capitalization.
For all networks that we have generated, the financial contagion starts from the single failure of a node, which affects neighboring nodes by default on its obligations in the
interbank lending market. Thus, this work focuses on the problem of credit risk, disregarding other equally important sources of contagion, as the risk of
adverse shocks reaching several institutions at the same time.

Some topological indices of systemic risk suggested in the literature (Cont et al. \cite{Rama2010}) are evaluated in order to verify its ability to explain
the impact of the failure of a node on the system.

The paper is structured as follows: section \ref{cap3} describes the model used for simulation of financial networks. We describe the methodology
used in the construction of connections between nodes, as well as the method for the simulation of their balance sheets. Section \ref{cap4} introduces
the method used to simulate financial contagion and presents the impact indices, by which we evaluate the nodes in respect to their default
effect. For each simulated network all nodes are analyzed individually. Section \ref{cap5} presents the results of the various
simulations performed by varying the parameters of the networks formation, the scale of the system (number of nodes) and the capital level. In section
\ref{cap6} we investigate the contribution of local characteristics of the networks in explaining the level of contagion. Section \ref{cap7} summarizes the
main conclusions.

\section{Generating scale free networks}\label{cap3}

In their study on scale free networks, Barabasi and Albert \cite{barabasi99} propose a preferential attachment mechanism to explain the emergence of the
power-law degree distribution in nondirected graphs. The resulting network exhibits a few nodes with very high connectivity (called hubs) and a large
number of nodes with a few or a single link. This is not really representative of real bank networks where there exist hubs but not so many single connected
nodes. Bollobás et al. \cite{conf/soda/BollobasBCR03} proposed a modification of the algorithm: a generalization for directed networks of the model
developed by Barabasi and Albert \cite{barabasi99}. In their model the network is formed by preferential attachment that depends on the distribution of
in-degree, $k_{in}$ and out-degree $k_{out}$.
This algorithm has the advantage of producing different exponents for the in and out degrees, which are necessary to reproduce the characteristics of
real networks (Allen and Babus \cite{Babus2009}). The following describes the steps for generating the network according to Bollobás et al.
\cite{conf/soda/BollobasBCR03}.

Let $\alpha$, $\beta$, $\gamma$, $\delta_{in}$ and $\delta_{out}$ be non-negative real numbers such that $\alpha+\beta+\gamma = 1$. Let $G_0$ be any initial
network\footnote{In this work the simulations performed use an initial network, $G_0$, consisted of 2 nodes, $0$ and $1$, connected by 2 directed links,
$0 \rightarrow 1$ e $1 \rightarrow 0$.}, and let $t_0$ be the number of links of $G_0$. At each step, $t$, starting with $t=t_0+1$, we add a new link to the network, so that in step
$t$ the network has $t$ links and a random number of nodes, $n(t)$. At each step the addition of the new link may be accompanied or not by adding new node,
according to the following method (Bollobás et al. \cite{conf/soda/BollobasBCR03}):

\begin{itemize}
 \item With probability $\alpha$, we add a new node $v$ with a link from $v$ to an existing node, $u$, selected with probability:
 \begin{equation}
  p(u=u_{i}) = \frac{k_{in}(u_i)+\delta_{in}}{t+n(t)\delta_{in}}
 \end{equation}
 \item With probability $\beta$, we select an existing node $v$ with probability:
 \begin{equation}
  p(v=v_{i})= \frac{k_{out}(v_{i})+\delta_{out}}{t+n(t)\delta_{out}}
 \end{equation}
 and add a link from $v$ to an existing node $u$, chosen with probability:
 \begin{equation}
  p(u=u_{i})= \frac{k_{in}(u_{i})+\delta_{in}}{t+n(t)\delta_{in}}
 \end{equation}
 \item With probability $\gamma$, we add a new node $u$ with a link from an existing node $v$ to $u$, where $v$ is selected with probability:
 \begin{equation}
  p(v=v_{i})= \frac{k_{out}(v_{i})+\delta_{out}}{t+n(t)\delta_{out}}
 \end{equation}
\end{itemize}

where $k_{in}(u_{i})$ is the in-degree of node $u_i$ and $k_{out}(v_{i})$ is the out-degree of node $v_i$.
Since the probability $\beta$ refers to the addition of a link without any creation of a node, increasing the value of $\beta$ implies increasing
the average network connectivity. In turn, the parameters $\alpha$ and $\gamma$ are related to the addition of new nodes while increasing the connectivity
of existing nodes.

As stated above, for interbank networks the in-degree of a node, $k_{in}$, represents its number of debtors, so large values of $\alpha$ tends to
generate nodes that concentrate many credits, i.e., the most connected nodes are large creditors. In a complementary way, large values of $\gamma$ (as
opposed to $\alpha$) tend to generate large debtors nodes. On the other hand, the parameters $\delta_{in}$ and $\delta_{out}$ represent weights distributed
among the nodes, causing everyone to have a chance of being selected in the attachment process. For example, with $\delta_{in} > 0$, even a node without
any {\it in} link can be selected to receive an {\it in} link with probability:

\begin{equation}
 p(u=u_{i}) = \frac{\delta_{in}}{t+n(t)\delta_{in}}
\end{equation}

The similar situation occurs when we have $\delta_{out} > 0$.

In their work, Bollobás et al. \cite{conf/soda/BollobasBCR03} show that, when the number of nodes goes to infinity and the connectivity grows, we have:

\begin{equation}
 p(k_{in}) \sim C_{IN}k_{in}^{-X_{IN}}
\end{equation}
\begin{equation}
 p(k_{out}) \sim C_{OUT}k_{out}^{-X_{OUT}}
\end{equation}
where:
\begin{equation}
 X_{IN} = 1 + \frac{1+\delta_{in}(\alpha+\gamma)}{\alpha+\beta}
\end{equation}
\begin{equation}
 X_{OUT} = 1 + \frac{1+\delta_{out}(\alpha+\gamma)}{\beta+\gamma}
\end{equation}

The limit $N \rightarrow \infty$ can obviously not be achieved, but the result is valid when the number of nodes grows and we take the
more connected ones, i.e., power laws for $k_{in}$ and $k_{out}$ will emerge in the tail of the distribution of large networks. Performing several simulations
using different parameter values we find that, as the network grows, the convergence to the limit values occurs from the left, i.e., from lower values than
predicted by the formulas of $X_{IN}$ and $X_{OUT}$.

We want to compare networks with different values for $X_{IN}$ and $X_{OUT}$ (featuring different concentrations of
$k_{in}$ and $k_{out}$), while keeping other characteristics similar, such as average connectivity and total concentration of links distribution.
We are particularly interested in networks whith values of $X_{IN}$ and $X_{OUT}$ around 2 and 3, in agreement with estimated empirical
values (for example, Boss et al. \cite{Boss2004}, Cont et al. \cite{Rama2010}, Soramäki et al. \cite{RePEc:fip:fednsr:243}). We restrict the degrees of
freedom of the model, imposing the following constraints on the parameters:

\begin{equation}\label{eq:restr1}
 \alpha+\gamma = 0,75
\end{equation}
\begin{equation}\label{eq:restr2}
 \delta_{in}+\delta_{out} = 4
\end{equation}

For $\alpha+\gamma = 0,75$, we have $\beta = 0,25$. Since $\beta$ is the probability of creating a link without addition of a new node, we can not use
a value of $\beta$ too small, otherwise we will have a network with very low average connectivity. On the other hand, we would like to have values of
$\alpha$ and $\gamma$ high enough for creating asymmetry between the distributions of in-degrees and out-degrees. The values $\alpha+\gamma = 0,75$ and
$\beta = 0,25$ satisfy those requirements. For example, the asymmetry between the exponents estimated $X_{IN}$ and $X_{OUT}$ for networks of 1000 nodes is
apparent if we use values $\alpha$ and $\gamma$ in the ratio of 1:3 (or 3:1), which implies, with $\alpha+\gamma = 0,75$, the values $\alpha = 0.1875$ and
$\beta = 0.5625$ (or $\alpha = 0.5625$ and $\beta = 0.1875$).

Also the differences between $\delta_{in}$ and $\delta_{out}$ cause asymmetry between the estimated exponents $X_{IN}$ and $X_{OUT}$. We use
$\delta_{in}$ and $\delta_{out}$ in the both ratios 1:3 and 3:1, in order to accentuate the asymmetry of the network.

In addition to the equations \ref{eq:restr1} and \ref{eq:restr2}, we will cover the spaces of parameters $\alpha \times \gamma$ e
$\delta_{out} \times \delta_{in}$ by varying the following radial lines:

\begin{equation}\label{eq:radiais}
 \alpha = \frac{\delta_{out}}{\delta_{in}}\gamma \; \rightarrow \;  \alpha = \frac{4-\delta_{in}}{\delta_{in}}\gamma
\end{equation}

\begin{figure}[htba]
    \centering
    \includegraphics[width=12cm]{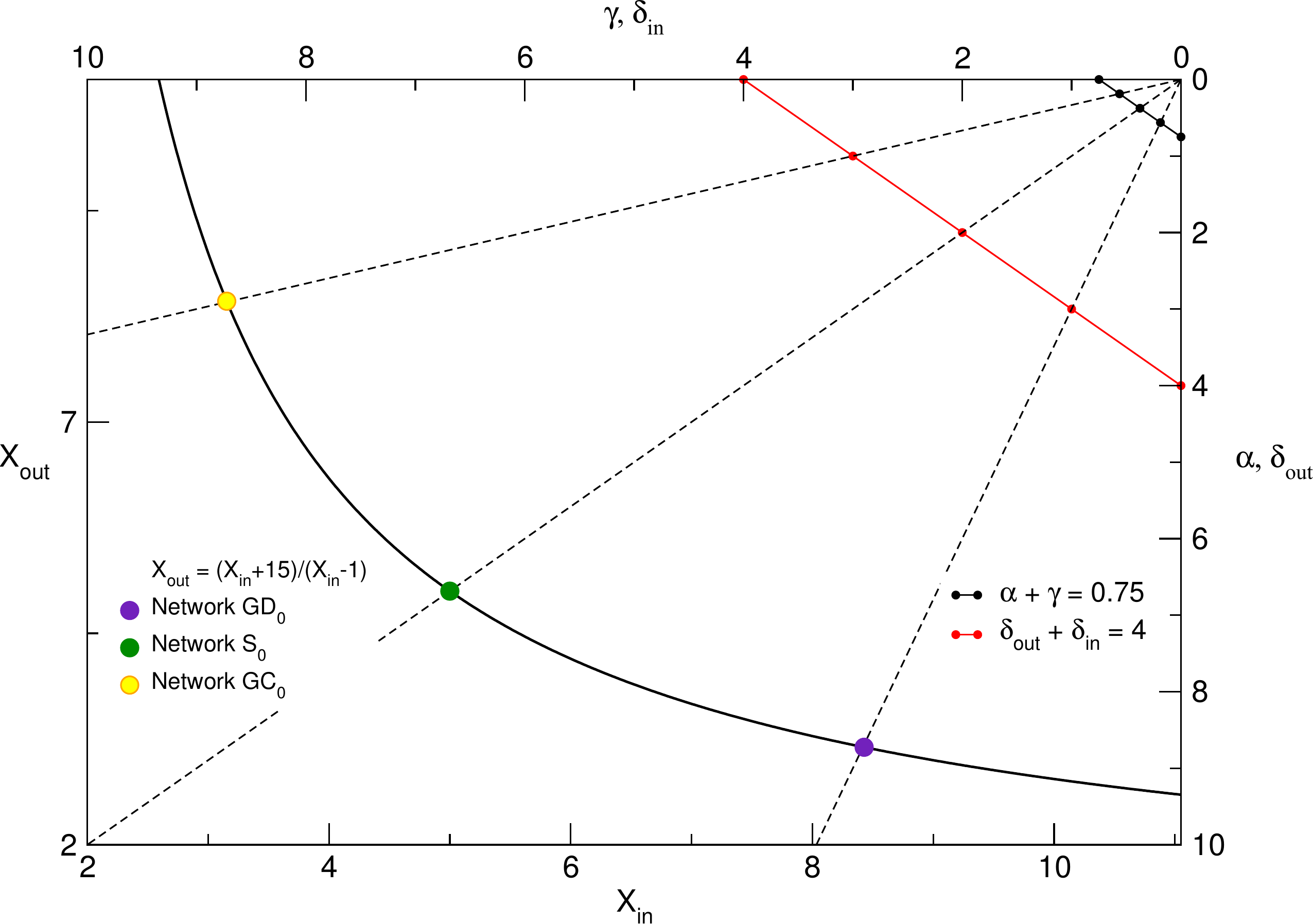}
    \caption[Espaço de parâmetros]{Parameter space and space of exponents: spaces $\alpha \times \gamma$ and $\delta_{out} \times \delta_{in}$
    are represented  with origin in the upper right corner and the space of exponents $X_{IN} \times X_{OUT}$ with origin at the bottom left.}
    \label{fig:parametros}
\end{figure}

The intersection points of equation \ref{eq:radiais} with the constraints \ref{eq:restr1} and \ref{eq:restr2} give us the set ($\alpha$, $\gamma$,
$\delta_{in}$, $\delta_{out}$), which in turn define pairs ($X_{IN}$, $X_{OUT}$) as shown in figure \ref{fig:parametros}.

Using the equations \ref{eq:restr1}, \ref{eq:restr2} and \ref{eq:radiais} we restate the parameters $\alpha$, $\beta$, $\gamma$ e $\delta_{out}$ as
functions of $\delta_{in}$:
\begin{equation}\label{eq:parametrica1}
 \alpha = \frac{12-3\delta_{in}}{16}
\end{equation}
\begin{equation}\label{eq:parametrica2}
 \gamma = \frac{3}{16}\delta_{in}
\end{equation}
\begin{equation}\label{eq:parametrica3}
 \delta_{out} = 4-\delta_{in}
\end{equation}
Replacing expressions \ref{eq:parametrica1}, \ref{eq:parametrica2} and \ref{eq:parametrica3} in the equations for $X_{IN}$ e $X_{OUT}$ respectively, we
obtain the two parametric equations:
\begin{equation}
 X_{IN} = 1+\frac{16+12\delta_{in}}{16-3\delta_{in}}
\end{equation}
\begin{equation}
 X_{OUT} = 1+\frac{68-9\delta_{in}}{4+3\delta_{in}}
\end{equation}
from which we finally have:
\begin{equation}\label{eq:curva}
 X_{OUT} = \frac{X_{IN}+15}{X_{IN}-1}
\end{equation}

The networks constructed using relationship \ref{eq:curva} are therefore generated through variation of a single degree of freedom, having similar
average connectivity and link concentration (limited by the constraints \ref{eq:restr1} and \ref{eq:restr2}), differing in the value of pairs
$(X_{IN}, X_{OUT})$. As we reported above, the exponent of a power law distribution reflects the concentration of the distribution: a smaller absolute value
of the exponent corresponds to a more concentrated distribution (Kunegis and Preusse\cite{kunegis:power-law}). Therefore, differences
between exponents $X_{IN}$ and $X_{OUT}$ represent differences between the concentrations in {\it in-out} degree distribution.

For our studies of contagion we selected three points on the curve in figure \ref{fig:parametros} (equation \ref{eq:curva}), representing three distinct
networks, denominated as $GD_0$, $S_0$ e $GC_0$. The network $GD_0$ is more concentrated in debtor side: with a higher concentration of debts than credits
it is generated so that the largest banks in the network are major debtors of the system. The network $GC_0$ has higher concentration of credits:
the biggest banks are major creditors of the network. The network $S_0$ corresponds to the symmetric case, in which the concentration of debts and
credits are similar. The limit values $(X_{IN}, X_{OUT})$ for these networks are shown in Table \ref{tab:1}.

\begin{table}[htbp]
    \caption[Three selected networks]{Limit values ??of the exponents $X_{IN}$ e $X_{OUT}$, average connectivity and Gini index for three selected networks.}
    \label{tab:1}
    \vspace{0.5em}
    \centering
    \begin{tabular}{l c c c c c c}
    \toprule
      & $X_{IN}$ & $X_{OUT}$ & $<k>$ & $G$ & $G_{in}$ & $G_{out}$\\
    \toprule
      {\bf Network $GD_0$} & 8.4286 & 3.1538 & 2.646 ($\pm 0.039$) & 0.457 ($\pm 0.006$) & 0.418 ($\pm 0.013$) & 0.746 ($\pm 0.009$)\\
      {\bf Network $S_0$} & 5.0000 & 5.0000 & 2.663 ($\pm 0.041$) & 0.429 ($\pm 0.006$) & 0.578 ($\pm 0.015$) & 0.576 ($\pm 0.012$)\\
      {\bf Network $GC_0$} & 3.1538 & 8.4286 & 2.652 ($\pm 0.028$) & 0.456 ($\pm 0.008$) & 0.748 ($\pm 0.011$) & 0.410 ($\pm 0.008$)\\
    \toprule
    \end{tabular}
\end{table}

\begin{figure}[htba]
    \centering
    \includegraphics[width=15cm]{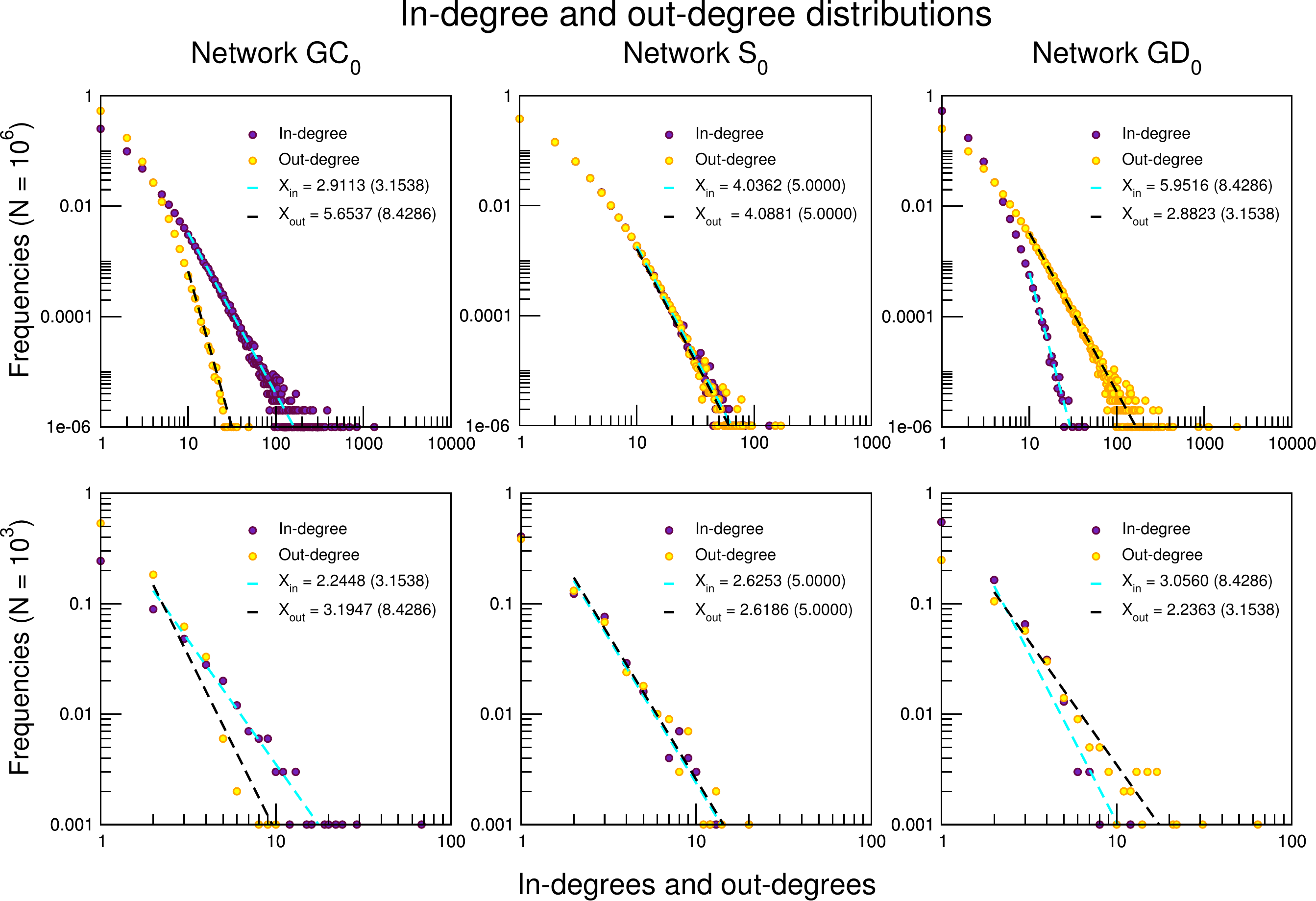}
    \caption[Selected networks]{Log-log plots for the in and out degrees distributions for three selected networks. First row: plots of networks
    $GC_0$, $S_0$ and $GD_0$ generated with $10^6$ nodes. In the second row: plots of the same networks generated with $10^3$ nodes. The values in parentheses next to estimated exponents refer to the limit values according to the model.}
    \label{fig:TresRedes}
\end{figure}

The choice of such values takes into account that we simulate small networks of 1,000 nodes (in agreement with real financial networks, for example,
Boss et al. \cite{Boss2004}, Soramäki et al. \cite{RePEc:fip:fednsr:243}, Mistrulli \cite{Mistrulli2007}), for which
the estimated exponents are below the limit values. Indeed, generating networks of 1,000 nodes with the selected parameters, we obtain exponents
between 2.2 and 3.2. Networks with these values of parameters are shown in figure \ref{fig:TresRedes}. The estimation of the exponents of the power law of
each distribution is done using the maximum likelihood estimator for discrete power laws, according to Clauset et al.\cite{Clauset2007}.

Table \ref{tab:1} also shows the values of the average connectivity for the generated networks, $<k>$, the Gini coefficient of distribution of links, $G$,
the Gini coefficient of the distribution of in-links (in-degree distribution), $G_{in}$, and the Gini coefficient of the distribution of out-links
(out-degree distribution). The values in the table are average values over 20 simulations for networks with 1,000 nodes.

\begin{figure}[htba]
    \centering
    \includegraphics[width=15cm]{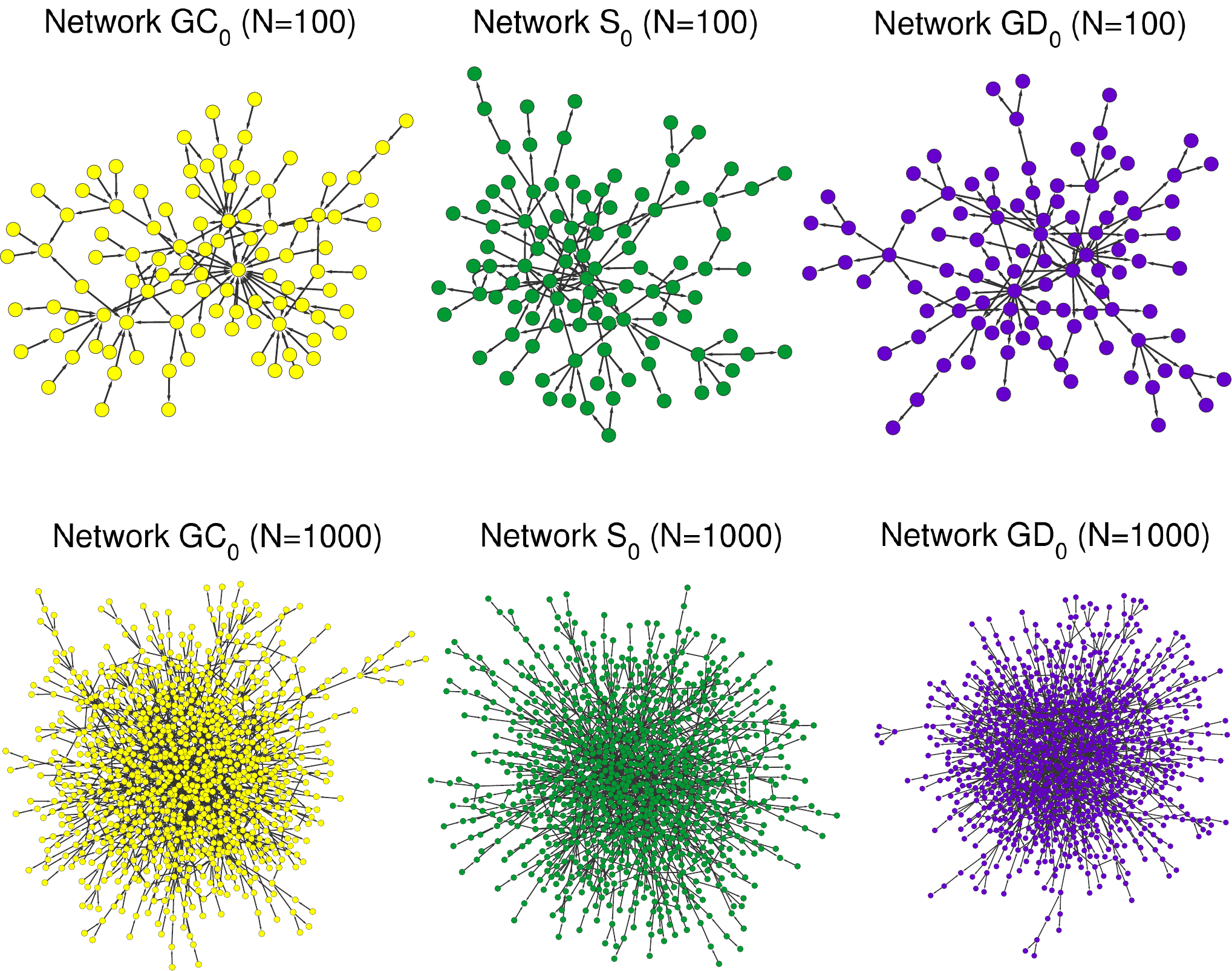}
    \caption[Visualization of selected networks: N=100 and N=1000]{Visualization of selected networks: N=100 and N=1000.}
    \label{fig:nets}
\end{figure}

For a better visualization of the networks, we have represented in figure \ref{fig:nets} examples of the networks $GC_0$, $S_0$ and $GD_0$, whith
number of nodes $N=100$ and $N=1000$. The smaller networks of 100 nodes provide better visualization of differences in the distribution of directed links.

In order to complete the information about an interbank network, it is necessary to assign weights to the links, since the weights represent the magnitude of
exposures between banks. The sum of in-degree weights of a bank, $i$, represents its applications in other institutions of the financial system
(loans to other banks), a variable that we define as {\it bank assets}, $BA_i$. The sum of out-degrees weights represents the total
obligations of $i$ to other financial institutions (loans from other banks), which we will call {\it bank liabilities}, $BL_i$. If there is a link from bank
$j$ to bank $i$, we define the exposure of bank $i$ to $j$ by $w_{ji}$, such that:

\begin{equation}
 BA_i = \sum_{j \in \{k_{in}^i\}}w_{ji}
\end{equation}
where $\{k_{in}^i\}$ is the set of banks having obligations to the bank $i$.
Similarly, if there is a link from bank $i$ to bank $j$, we define the obligation of bank $i$ to bank $j$ by $w_{ij}$, such that:
\begin{equation}
 BL_i = \sum_{j \in \{k_{out}^i\}}w_{ij}
\end{equation}
where $\{k_{out}^i\}$ is the set of banks for which bank $i$ has obligations to pay.

In a study on the Brazilian interbank network, Cont et al.\cite{Rama2010} highlight the non-linear positive relationship between link weights and
connectivity of nodes in line with the widespread notion that the size of balance sheets and connectivity of banks are positively related (Arinaminpathy et
al.\cite{RePEc:boe:boeewp:0465}).
From this assumption and in order to compute the weight of each of the links of a node, we define the following expression for the weight of the links:

\begin{itemize}
 \item For a link from $i$ to $j$:
 \begin{equation}\label{eq:peso1}
  w_{ij} = \frac{k_{out}^{i} \cdot k_{in}^{j}}{k_{out}^{max} \cdot k_{in}^{max}}
 \end{equation}
 \item For a link into $i$ from $j$:
 \begin{equation}\label{eq:peso2}
  w_{ji} = \frac{k_{in}^{i} \cdot k_{out}^{j}}{k_{in}^{max} \cdot k_{out}^{max}}
 \end{equation}
\end{itemize}
In equations \ref{eq:peso1} and \ref{eq:peso2} $k_{in}^{max}$ and $k_{out}^{max}$ denote the maximum values of $k_{in}$ and $k_{out}$ found in the network.
The values $w_{ij}$ form matrix $n \times n$ of mutual exposures, $W$, where each element $w_{ij} \in W$ represents the exposure of the bank $j$ to the
bank $i$. Note that the level of mutual exposures between any two banks is closely related to the local structure of the network.

Once established the values of bank assets and bank liabilities, $BA_i$ and $BL_i$, we define the other elements of the balance sheet: nonbank assets,
nonbank liabilities and equity.

\begin{enumerate}
 \item{Nonbank assets, $NBA$:}
 refer to all applications except interbank ones. They include loans to non-financial firms and households.
 \item{Nonbank liabilities, $NBL$:}
  refer to funding from outside the system, mostly represented by deposits of non-financial agents.
 \item{Equity, $E$:}
  represents shareholders' funds invested in the bank, i.e., the partners' capital.
\end{enumerate}

The balance sheets are represented in simplified form, in which we do not take into account elements such as different maturities and differences in
assets liquidity and risk.

For each bank, $i$, the balance sheet obey the identity that total assets equals total liabilities:
\begin{equation}\label{eq:identidade}
 BA_{i}+NBA_{i}=BL_{i}+NBL_{i}+E_{i}
\end{equation}

Reflecting the minimum capital regulations of Basel Accords we set equity of each bank as a proportion of its assets:

\begin{equation}\label{eq:PL}
 E_i = \lambda_i (BA_i+NBA_i)
\end{equation}
where $\lambda_i$ represents the capital/assets ratio.

For the simulations in this work we will adopt three values of capital/assets ratio: the undercapitalized case, with $\lambda=0.01$, and values
$\lambda=0.05$ and $\lambda=0.1$, consistent with the empirical values observed (IMF \cite{IMF}). For each bank the capital/asset ratio is extracted from a
normal distribution $\lambda_i \sim N(\lambda, \sigma)$ subject to the constraint $\lambda_i > \lambda \nonumber$, i.e., $\sigma$ is a stochastic
positive deviation from the minimum $\lambda$, characterizing the heterogeneity of banks as regard to capitalization. The simulations are performed
using $\sigma=0.01$.

To represent the ratio of nonbank assets to total assets, we introduce the following relation that defines the nonbank assets for each bank, $i$, as:

\begin{equation}\label{eq:ANB}
 NBA_{i} = \xi(BA_{i}+BL_{i})
\end{equation}

Defined this way, nonbank assets are a function of bank connectivity (via $BA_{i}$ and $BL_{i}$), maintaining consistency with the
assumption that the balance size is related to connectivity. Let's use $\xi$ as calibration factor to control the $NBA_{i}$ to total assets ratio.

The identities \ref{eq:identidade}, \ref{eq:PL} and \ref{eq:ANB} form a system of equations by which the value of $NBL_{i}$ can be determined:
\begin{equation}\label{eq:PNB}
 NBL_{i} = (1-\lambda_{i})(1+\xi)BA_{i} + [(1-\lambda_{i})\xi-1]BL_{i}
\end{equation}

Thus, the nonbank assets to total assets ratio and the nonbank liabilities to total liabilities ratio are:

\begin{equation}\label{eq:ANBA}
 \frac{NBA_{i}}{A_{i}}=\frac{\xi (BA_{i}+BL_{i}) }{(\xi+1)BA_{i}+\xi BL_{i}}
\end{equation}
\begin{equation}\label{eq:PNBP}
 \frac{NBL_{i}}{L_{i}}=\frac{(1-\lambda_{i})(1+\xi)BA_{i}+[(1-\lambda_{i})\xi-1]BL_{i}}{(1-\lambda_{i})(1+\xi)BA_{i}+(1-\lambda_{i})\xi BL_{i}}
\end{equation}

As $BA_{i}$ e $BL_{i}$ are random values (depending on network formation process) we evaluate the equations \ref{eq:ANBA} and \ref{eq:PNBP} in
three possible limits:

\begin{enumerate}
 \item{$BA_{i} \ll BL_{i}$:}
 \begin{equation}
  \frac{NBA_{i}}{A_{i}} \rightarrow 1
 \end{equation}
 \begin{equation}
  \frac{NBL_{i}}{L_{i}} \rightarrow \frac{(1-\lambda_{i})\xi-1}{(1-\lambda_{i})\xi}
 \end{equation}
 \item{$BA_{i} = BL_{i}$:}
  \begin{equation}
  \frac{NBA_{i}}{A_{i}} = \frac{2\xi}{2\xi+1}
 \end{equation}
 \begin{equation}
  \frac{NBL_{i}}{L_{i}} = \frac{2\xi(1-\lambda_{i})-\lambda_{i}}{2\xi(1-\lambda_{i})-\lambda_{i}+1}
 \end{equation}
 \item{$BA_{i} \gg BL_{i}$:}
  \begin{equation}
  \frac{NBA_{i}}{A_{i}} \rightarrow \frac{\xi}{\xi+1}
  \end{equation}
  \begin{equation}
  \frac{NBL_{i}}{L_{i}} \rightarrow 1
  \end{equation}
\end{enumerate}

\begin{figure}[htbp]
    \centering
    \includegraphics[width=15cm]{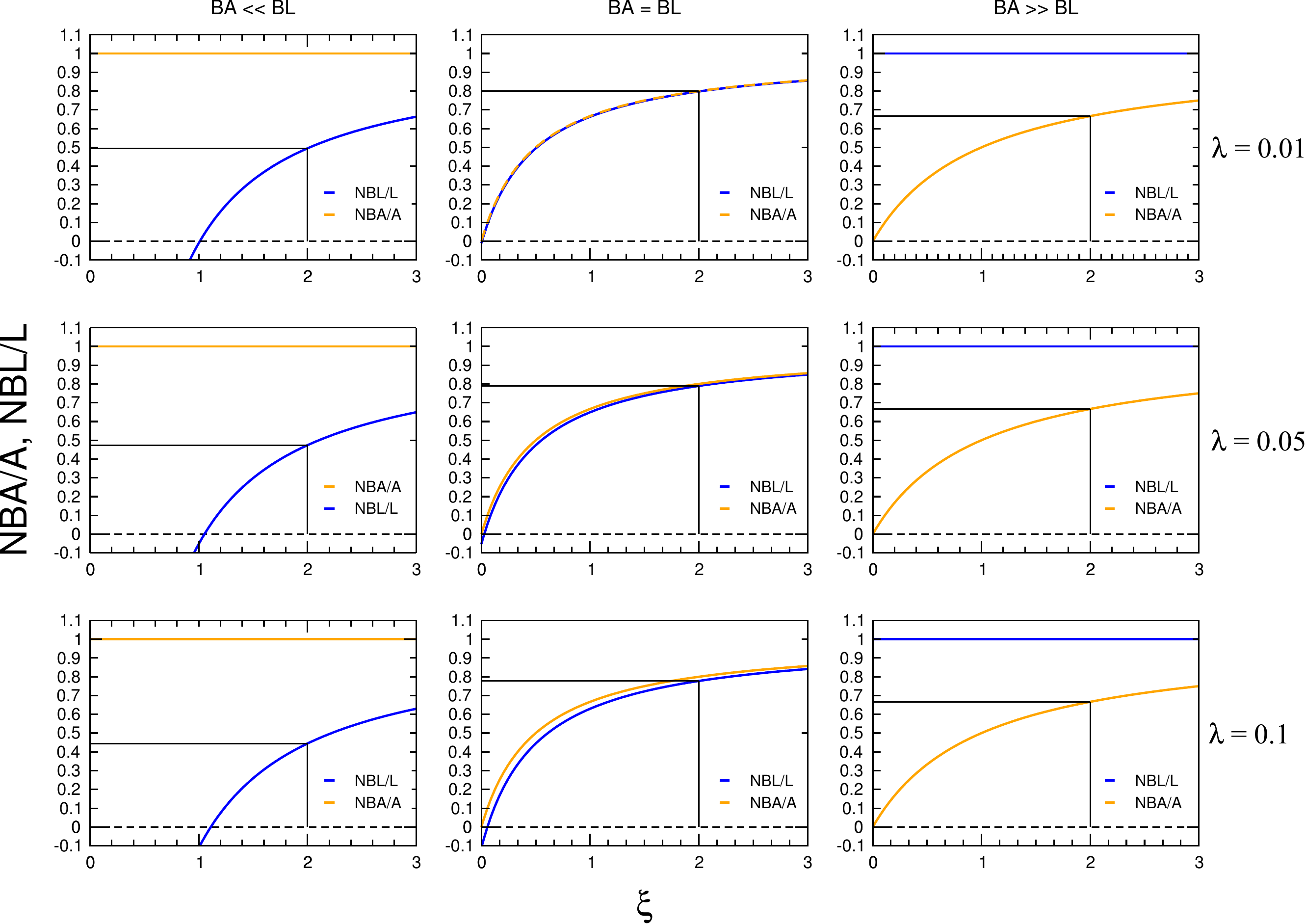}
    \caption[Ratios $NBA/A$ e $NBL/L$]{Ratios $NBA/A$ and $NBL/L$ as functions of $\xi$ for the three values of $\lambda$.}
    \label{fig:BnB}
\end{figure}

Figure \ref{fig:BnB} shows the ratios \ref{eq:ANBA} and \ref{eq:PNBP} for different values of $\xi$ and for the three values of $\lambda$. For the
simulations in this work we fix $\xi = 2$ in order to obtain balance sheets in which nonbank assets and nonbank liabilities represent on average more than
50\% of total assets and liabilities.

From the method described in this section we are able to represent the balance sheet of each bank by using only information from the network and the
parameters $\lambda$ and $\xi$. In the following section we will describe the cascade of failures following the initial default of one bank of the network.

\section{Contagion in interbank networks: default cascade and default impact}\label{cap4}

In this section we present the methodology used to evaluate the propagation of losses in the interbank network. We simulate the insolvency of a single bank,
exposed to an external shock represented by the total loss of value of its nonbank assets. Each bank is tested independently and the impact of its default
on the system evaluated.

In a hypothetical scenario a bank, $i$, becomes insolvent, being unable to completely fulfill its obligations. If at time $t$, bank $j$ realize that
its counterparty $i$ is unable to repay its interbank liability $w_{ij}$ in full, then bank $j$ must reevaluate its application in bank $i$, from $w_{ij}$
to $w'_{ij}$: $(w'_{ij}-w_{ij})<0$. This process adversely affect the capital of $j$, since variation $(w'_{ij}-w_{ij})$ is incorporated as a loss. It
happens that the smaller value, $w'_{ij}$, the defaulting bank $i$ can effective afford, depends on the financial conditions of other banks, banks for
which $i$ had granted loans. Any further failure reduces the value of assets, increasing the losses of banks that have already defaulted.

In their work Eisenberg and Noe \cite{Eisenberg.Noe2001SystemicRiskin} study the problem of calculating the values $w'_{ij}$ that banks would be able to pay at the
time of settlement of its multilateral obligations. Given the array of mutual exposures, $W$, the problem is to determine the vector of payments,
$p= (p_1, p_2,...,p_n)$, where:

\begin{equation}
 p_i = \sum_{j=1}^{n}w'_{ij}
\end{equation}

The authors show that under mild regularity conditions, there is a unique payment vector that settle the system, and develop an iterative
algorithm to solve the problem. In the context of our work, where we have a single node initially insolvent, the algorithm can be described as follows:
\begin{enumerate}
 \item Compute the losses to all banks resulting from the failure of bank $i$ assuming that all other banks are able to repay their liabilities. Stop
 if no other bank fails, otherwise:
 \item Let $j$ denote the bank or group of banks whose losses exceed their equity. Compute the losses to all banks resulting from the failure of banks $i$
 and $j$. Repeat step 2 until no further bank fails\footnote{For more details about the algorithm see the original paper of Eisenberg and Noe
 \cite{Eisenberg.Noe2001SystemicRiskin}.}.
\end{enumerate}

The algorithm described above allows us to calculate two important measures to assess the impact of a bank failure on the network:
the {\it Default Impact} and {\it Default Cascade}. For a bank $i$, the {\it Default Impact}, $DI_i$ refers to the reduction in total assets of the
financial system as a result of losses incurred via contagion, as a proportion of total initial assets. If we denote the total assets of the system at the
initial time as $A_0$ and at final time (after the external shock) as $A_t$, the {\it Default Impact} is given by:

\begin{equation}
 DI_i = \frac{A_0-A_t-NBA_i}{A_0}
\end{equation}\label{eq:impacto}

The equation (36) excludes from the computation of $DI_i$ the value of {\it Initial Impact} ($NBA_i$), representing only the losses due to contagion. The
{\it Total Impact}, $TI_i$, is simply the sum of the {\it Initial Impact} and the {\it Default Impact}:

\begin{equation}
 TI_i = \frac{NBA_i}{A_0} + \frac{A_0-A_t-NBA_i}{A_0}
\end{equation}\label{eq:IT_i}

The measure {\it Default Cascade}, $DC_i$, refers to the number of insolvent banks due to the failure of bank $i$,  as a proportion of the total number
of banks of the network. Both the {\it Default Impact} and {\it Default Cascade} of a bank reveal how the network would be affected by its failure,
taking into account only the direct effects of loss propagation through interbank exposures. Although the {\it Default Impact} seems the most relevant
measure, since it refers to the magnitude of impact in terms of asset value, equally important is the {\it Default Cascade}, since the way a crisis is felt
by economic agents is also related to the number of affected  banks.

To evaluate the sensitivity of contagion to local measures of connectivity and concentration, Cont et al. \cite{Rama2010} propose the indices
{\it Counterparty susceptibility}, $CS(i)$, and {\it Local network frailty}, $f(i)$. Below we reproduce the definitions of the authors \cite{Rama2010}.

{\bf Definition} ({\it Counterparty susceptibility}): The counterparty susceptibility $CS(i)$ of a node $i$ is the maximal (relative) exposure to node $i$
of its counterparties:

\begin{displaymath}
 CS(i) = \max_j \frac{w_{ij}}{E_j}
\end{displaymath}

where $E_j$ is the total exposures of counterparty $j$.

$CS(i)$ is a measure of the maximal vulnerability of creditors of $i$ to the default of $i$.

{\bf Definition} ({\it Local network frailty}): The local network frailty $f(i)$ at node $i$ is defined as the maximum exposure to node $i$ of its
counterparties $j$ (in \% of capital of $j$), weighted by the size of their interbank liability:

\begin{displaymath}
 f(i) = \max_j \frac{w_{ij}}{E_j}BL_j
\end{displaymath}

As well as the measure $CS_i$, the {\it Local fragility network} is a measure of the vulnerability of creditors, but it also reflects the sensitivity
of indirect lenders of $i$, represented by the liabilities of counterparty $j$, $BL_j$.

For the simulations implemented in this paper we calculate the indices $CS(i)$ and $f(i)$ in order to verify the relationship between these topological
characteristics and the {\it default impact } and {\it defaults cascade} of the nodes.

\section{Results}\label{cap5}

In this section we present the results obtained in contagion simulations for networks produced according to the methodology presented in section \ref{cap3}.
We have three principal networks as already defined: network $GD_0$ presents a higher concentration of debts than credits, network $GC_0$ has higher
concentration of credits and $S_0$ is symmetric, generated with equal concentrations.

\subsection{Default Impact and Default Cascade}\label{redes0}

For each set of parameters that defines a network category ($GD_0$, $S_0$ and $GC_0$) we implemented 20 simulations, so that the analysis is based
on 20 realizations of networks of type $GD_0$, 20 realizations of $S_0$ networks and 20 realizations of $GC_0$ networks. For each generated network
and for each bank, $i$, the {\it Default Impact}, $DI_i$ and {\it Default Cascade}, $DC_i$, were calculated. The results presented in this section are for
networks with 1000 nodes, with capital level of $\lambda=0.05$.

\begin{figure}[htba]
    \centering
    \includegraphics[width=12cm]{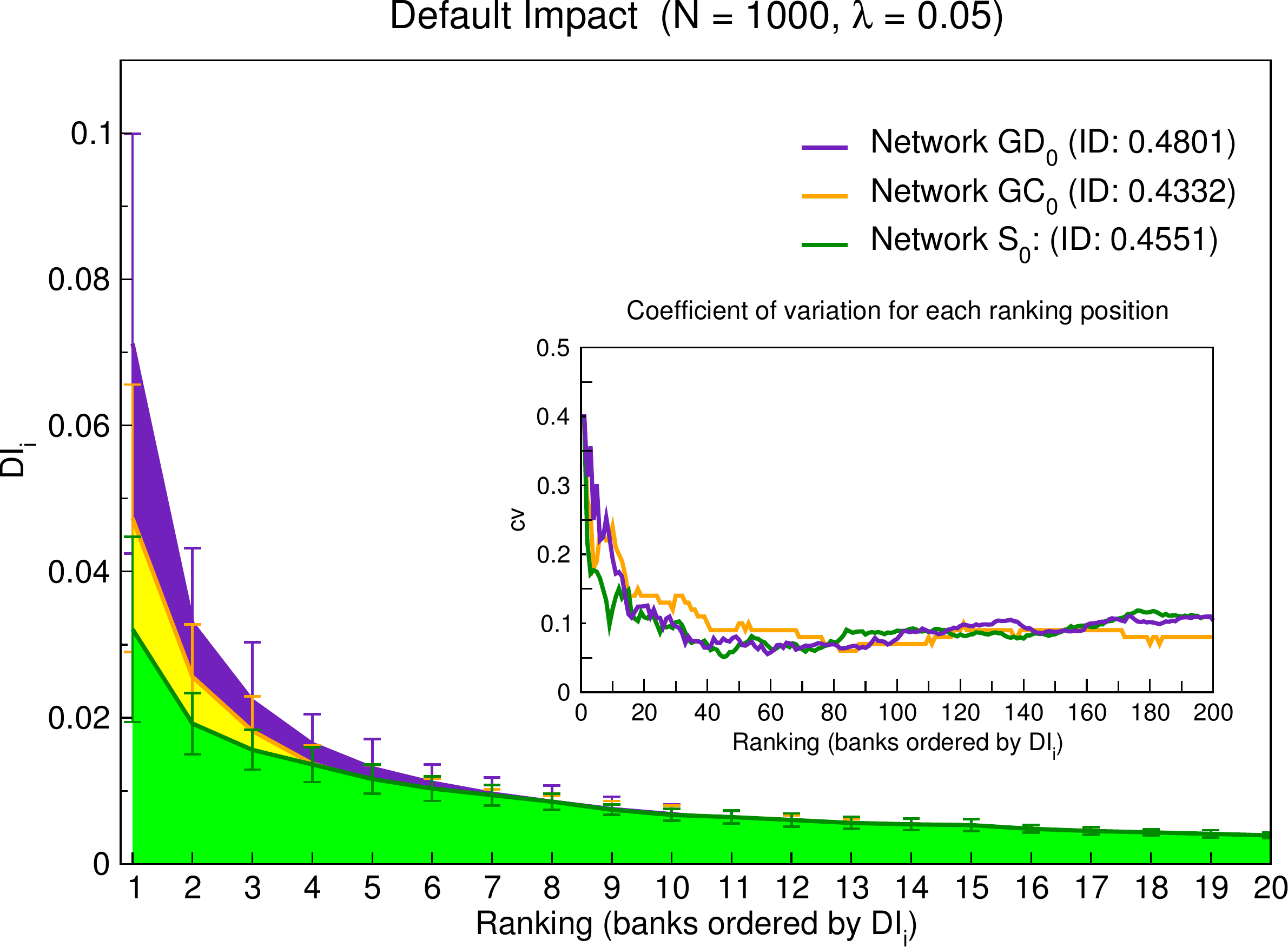}
    \caption[Ranking of banks in decreasing order of {\it Default Impact}, $DI_i$]{Ranking of banks in decreasing order of $DI_i$: The figure shows the
    values ??for the first banks causing major impact. Recall that, for each bank, $i$, the {\it Default Impact}, $DI_i$ refers to the losses suffered by
    the system via contagion (from the default of $i$) as a proportion of total assets of the network. The inset graph presents the values of the
    coefficient of variation ($cv = \sigma/\mu)$ at each ranking position for the 200 banks with greater impact.}
    \label{fig:ID_0}
\end{figure}

Figure \ref{fig:ID_0} shows the ranking of banks for the three networks, in decreasing order of $DI_i$. The values shown are average values
for each ranking position, for example, for each network type the greater {\it Default Impact} (first ranking position) is the average of greater
impacts for 20 simulations. Equivalently, the subsequent positions of the ranking are average values.

As we can see, the difference between the three types of networks is more pronounced in the first ranking positions, although these
positions show greater dispersion around the mean value: in the first position the standard deviation is, for the three categories of networks, around 40\%
of the average, as shown in the inset graph of Figure \ref{fig:ID_0}.

One can consider the area under the ranking curve, which corresponds to the sum of individual impacts, as a measure of the network systemic risk.
We then have for each network an aggregate measure, $DI$, given by:

\begin{equation}
 DI = \sum_{i=1}^{n}DI_i
\end{equation}

The measure $DI$ corresponds to a measure of central tendency: in fact, if $DI$ is divided by $N$ (number of nodes) we have the average value of individual
default impact. Ordering the three networks by the aggregate index, $DI$, we have $GD_0$ network with greater impact
($DI$=0.48), followed by $S_0$ ($DI$=0.46), and finally the $GC_0$ network ($DI$=0.43) (see figure \ref{fig:ID_0}). The ordering follows the direction of
increasing concentration of debt links, from the highest concentration in the $GD_0$ network, then a moderate concentration in $S_0$, until the smallest
concentration in $GC_0$.

Otherwise, if the three networks are evaluated according to their principal banks (first banks in the ranking), the network $GD_0$ remains in the
first position of impact, but the other two switch positions: banks of network $GC_0$ with higher $DI_i$ presents greater effect over the system than
the big banks of network $S_0$. This fact can be explained by differences in the asset concentrations of networks $GD_0$, $GC_0$ and $S_0$. It happens that
banks with large balance sheets impact more strongly the system and although we have constructed the networks in a way that they present similar
concentrations of connectivity, interbank exposures as defined by the model accentuates assets concentrations, also increasing the differences between
them. In fact, the Gini coefficient for asset concentration is 0.83 for the network $GC_0$, 0.80 for $GD_0$ and 0.78 for $S_0$\footnote{The concentration
of assets in real networks is also quite high, as reported in the literature. For example, \cite{Elsinger.Lehar.ea2006RiskAssessmentBanking} report a Gini
of 0.88 for the Austrian network in 2002 and \cite{Ennis2001} reports a Gini of 0.90 for the United States in 2000.}. With the highest Gini, the network
$GC_0$ has a large bank whose total asset is 122 times greater than the average assets of the network, while the largest bank in the symmetric network $S_0$
owns a total asset 55 times greater than average.

The comparison between networks requires caution, since an estimation of actual losses would involve a study of stress scenarios leading to
default of a node or a set of nodes of the system.

\begin{figure}[htba]
    \centering
    \includegraphics[width=12cm]{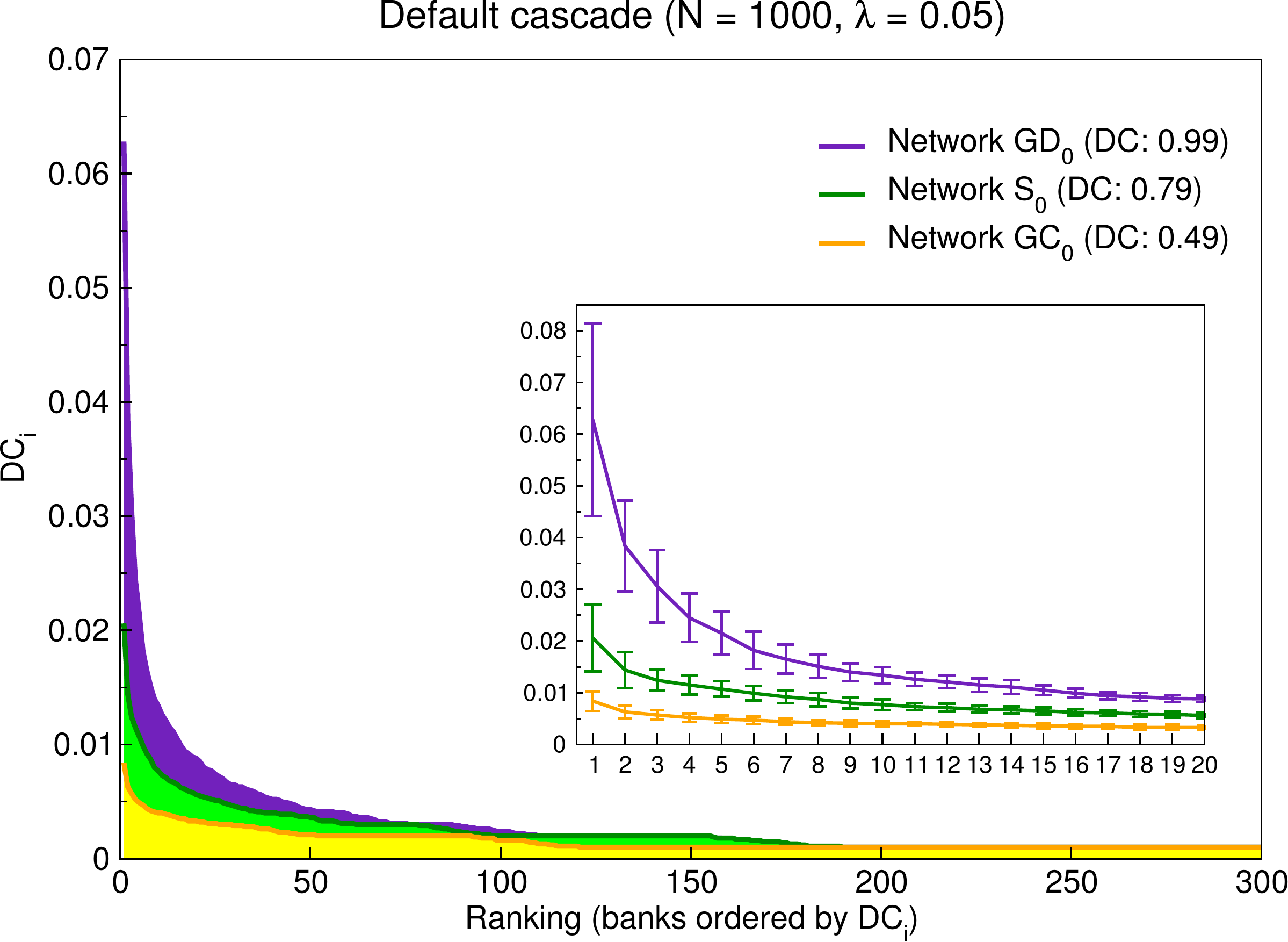}
    \caption[Ranking of banks in decreasing order of {\it Default Cascade}, $DC_i$]{Ranking of banks in decreasing order of $DC_i$: The figure shows the
    values ??for the 300 banks causing greater cascades. As defined earlier, the Cascade Default, $CD_i$, refers to the number of insolvent banks due to the
    initial failure of bank $i$ as a proportion of total number of banks in the network. The inset graph shows in detail the top 20 banks.}
    \label{fig:CD_0}
\end{figure}

Comparison is more straightforward when the networks are evaluated by the {\it default cascade} of their nodes, because in this case the difference is
more pronounced, as shown in figure \ref{fig:CD_0}. For {\it default cascade} we also define as aggregate measure the area under the ranking curve:

\begin{equation}
 DC = \sum_{i=1}^{n}DC_i
\end{equation}

Here the ordering of networks is clear: the network $GD_0$ has greater potential to generate contagion in case of default of its nodes.
Secondly we have the symmetric network $S_0$, and finally the network $GC_0$. The size of the balance sheet has less influence on the {\it Default Cascade}
than over the {\it Default Impact}, and the effect of concentrations of debt and credit links becomes more apparent. In fact, as expected, the
{\it Default Cascade} increase towards increased concentration of debts. Although having similar sizes, the bank that leads to higher cascade in network
$GD_0$ reaches about 6\% nodes of the network, compared with less than 1\% for $GC_0$ network.

\subsection{The effect of network size}\label{escala}

To evaluate the effect of scale on the indices of contagion, we vary the network size for the three kinds of network analyzed, producing networks with
number of nodes $N=500$, $N=1000$, $N=5000$ e $N=10000$. The parameters are the same as those used in the previous section and therefore the average
connectivity remains the same. The variation of size causes no change in the $DI$ and $DC$ measures.
This feature implies that the average values of individual impacts, $DI_i$ and  default cascade $DC_i$, decrease when network increases, minimizing the
importance of each node as the number of nodes increases. Thus, we conclude that larger networks exhibit nodes with less potential for contagion, situation
already observed in previous simulation works (Cont and Moussa \cite{Rama2010b}).

\begin{figure}[htba]
    \centering
    \includegraphics[width=\textwidth]{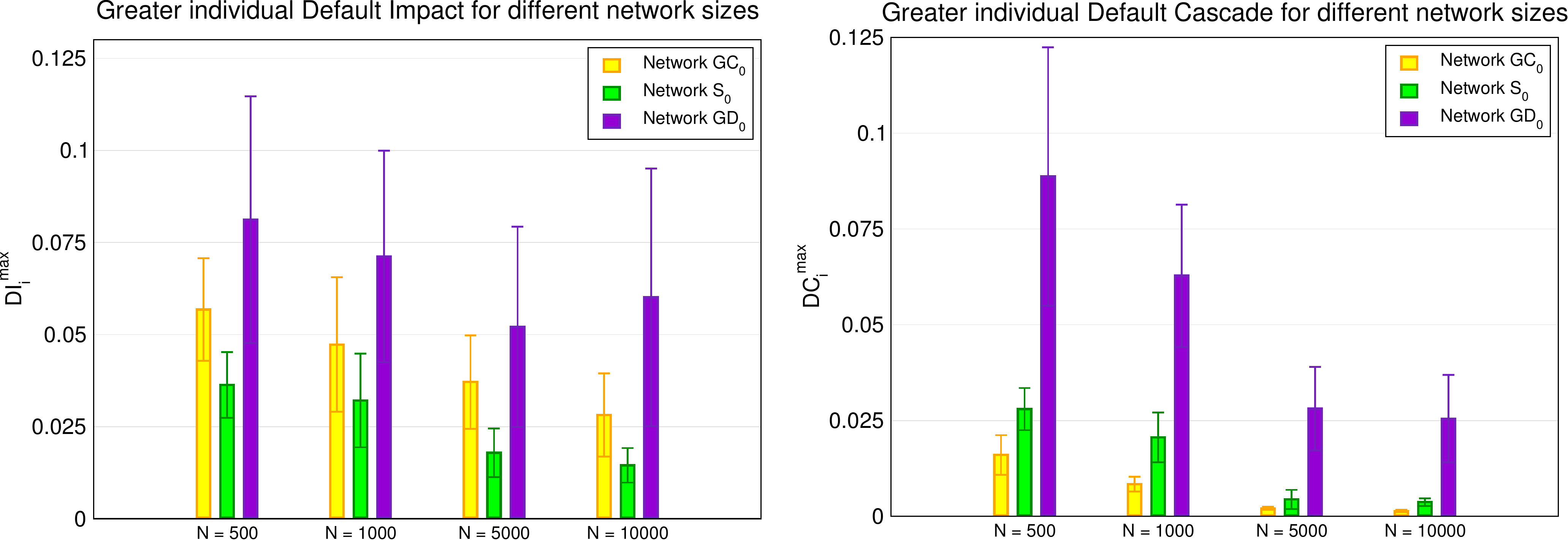}
    \caption[Effect of the network size on $DI_i$ and $DC_i$ of nodes of greater impact]{Left: values of {\it Default Impact} of the first ranking position
    (bank with larger $DI_i$), $DI_i^{max}$, for different network sizes. Right: values of {\it Default Cascade} of the first ranking position
    (bank with larger $DC_i$), $DC_i^{max}$, for different network sizes.}
    \label{fig:EscalaBIG_di_dc}
\end{figure}

Figure \ref{fig:EscalaBIG_di_dc} presents the values of the first position in the rankings of $DI_i$ and $DC_i$ for the three categories of network and
different sizes, showing the declining impact of individual nodes.

\subsection{Effect of capital level}\label{capital}

\begin{figure}[htba]
    \centering
    \includegraphics[width=\textwidth]{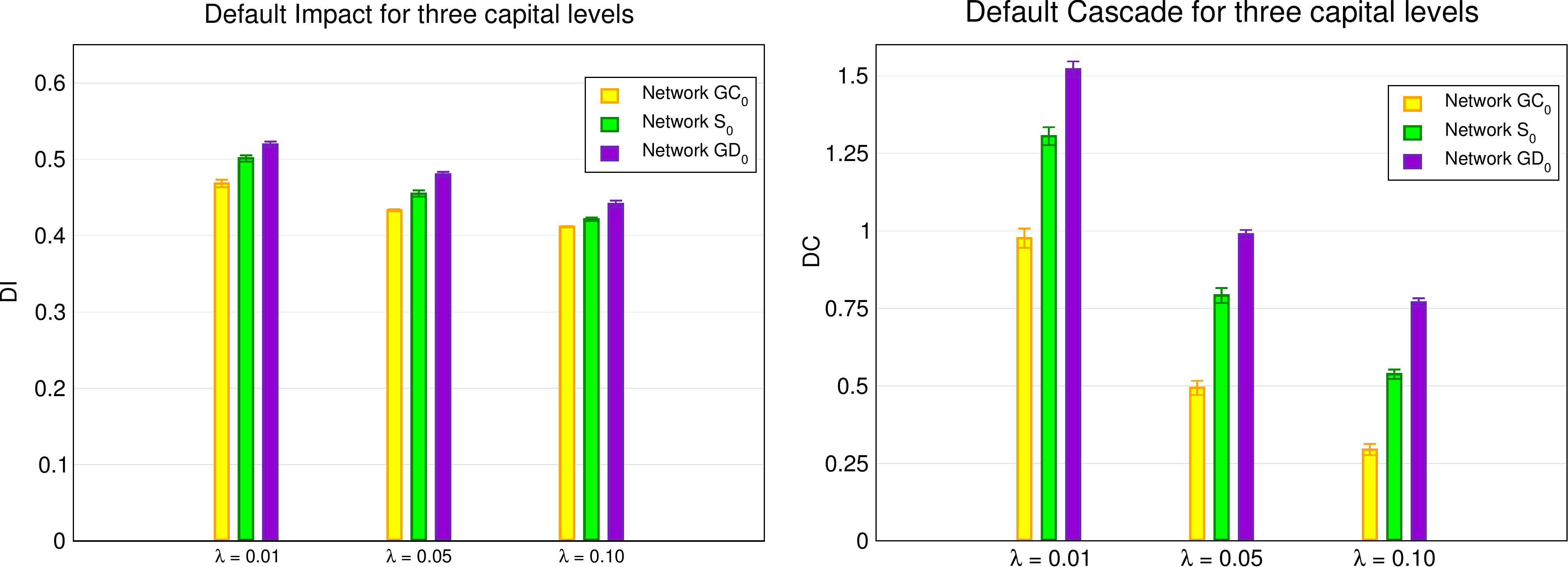}
    \caption[Effect of capital level on aggregate measures, $DI$ and $DC$]{Left: values of $DI$ in the three categories of network for $\lambda=0.01$,
    $\lambda=0.05$ and $\lambda=0.10$. Right: values of $DC$ in the three classes of network for $\lambda=0.01$, $\lambda=0.05$ e $\lambda=0.10$.}
    \label{fig:lambda}
\end{figure}

We evaluated the effect of a change in the capital level, $\lambda$ on the {\it Default Impact} and {\it Default Cascade}. The aggregate measures,
$DI_i$ and $DC_i$ are shown in figure \ref{fig:lambda}.

As expected, the increase of capital is responsible for a decrease in contagion for all network categories. Obviously a decrease in capital has the reverse
effect. Testing the network for lower values of $\lambda$ gives us an idea of possible amplified contagion in the event of macroeconomic stress, when much
of the network may become less capitalized. As one can see in figure \ref{fig:lambda}, the variation of capital level has more effect on the
{\it Default Cascade} than on {\it Default Impact}.

\subsection{Effect of connectivity and links concentration}\label{conectividade}

Previous studies that simulate contagion in different network topologies address the question of the influence of connectivity on the propagation of losses.
Conclusions differ, depending strongly on the structure of the networks used in each work.

It should be considered that if we simply increase network connectivity through addition of new random links, the distribution of connectivity turns less
concentrated, which also affects network performance in propagation of losses. We would like to separate these two effects: the effect of connectivity
variation and changes in concentration. Here the term concentration refers to the distribution of links, in the sense that a more concentrated networks is a
network where few nodes have many links (hubs), while the majority of nodes possess just a few. A less concentrated networks presents a more egalitarian
distribution of links.

In an attempt to understand how these two features act on financial contagion, we have tested new types of network which are variants of the three networks
previously evaluated, with different levels of connectivity and concentration. Thus, in addition to the original type, which we have named type $0$ and
consist of $GD_0$, $S_0$ and $GC_0$, we have implemented 4 new types, each one detailed in the following sections. The type 1 networks have greater
connectivity and are more concentrated than the original. Type 2 networks have greater connectivity and similar concentration to type 0. Type
3 networks have greater connectivity than type 0 but are less concentrated. Finally, type 4 networks maintain the same connectivity as the original
type, but with lower concentration. Table \ref{tab:2} shows the parameters used in the construction of networks of type 0, 1, 2, 3 and 4.

 \begin{table}[htbp]
     \caption[Parameter for the networks $GD$, $S$ e $GC$ (tipos 0 a 4)]{Parameter values ??used in the construction of networks of type 0, 1, 2, 3 and 4.}
    \label{tab:2}
    \centering
    \begin{tabular}{l c c c c c}
    \toprule
      & $\alpha$ & $\beta$ & $\gamma$ & $\delta_{in}$ & $\delta_{out}$\\
    \toprule
      {\bf $GC_0$} & 0.5625 & 0.2500 & 0.1875 & 1.00 & 3.00\\
      {\bf $S_0$} & 0.3750 & 0.2500 & 0.3750 & 2.00 & 2.00\\
      {\bf $GD_0$} & 0.1875 & 0.2500 & 0.5625 & 3.00 & 1.00\\[2.0mm]
      {\bf $GC_1$} & 0.1875 & 0.7500 & 0.0625 & 1.00 & 3.00\\
      {\bf $S_1$} & 0.1250 & 0.7500 & 0.1250 & 2.00 & 2.00\\
      {\bf $GD_1$} & 0.0625 & 0.7500 & 0.1875 & 3.00 & 1.00\\[2.0mm]
      {\bf $GC_2$} & 0.1875 & 0.7500 & 0.0625 & 25.00 & 75.00\\
      {\bf $S_2$} & 0.1250 & 0.7500 & 0.1250 & 50.00 & 50.00\\
      {\bf $GD_2$} & 0.0625 & 0.7500 & 0.1875 & 75.00 & 25.00\\[2.0mm]
      {\bf $GC_3$} & 0.5625 & 0.2500 & 0.1875 & 1.00 & 3.00\\
      {\bf $S_3$} & 0.3750 & 0.2500 & 0.3750 & 2.00 & 2.00\\
      {\bf $GD_3$} & 0.1875 & 0.2500 & 0.5625 & 3.00 & 1.00\\[2.0mm]
      {\bf $GC_4$} & 0.5625 & 0.2500 & 0.1875 & 10.00 & 30.00\\
      {\bf $S_4$} & 0.3750 & 0.2500 & 0.3750 & 20.00 & 20.00\\
      {\bf $GD_4$} & 0.1875 & 0.2500 & 0.5625 & 30.00 & 10.00\\
    \toprule
    \end{tabular}
\end{table}

\subsubsection{Greater connectivity and greater links concentration}

In this case we have the networks $GD_1$, $S_1$ and $GC_1$, constructed so that they have greater connectivity and are more concentrated than the
original ones. This is done by increasing $\beta$ from 0.25 to 0.75. We maintain the ratio 1:3 between $\alpha$ and $\gamma$ and the same
values of $\delta_{in}$ e $\delta_{out}$. As the probability $\beta$ refers to addition of new link without creation of new node, increasing $\beta$ implies
increasing the average network connectivity.
As the connection mechanism is preferential attachment, ie nodes that receive links are chosen with probability that is proportional to its previous
connectivity, an increase of $\beta$ also increases links concentration in higher connected nodes, so that the Gini coefficient for
links distribution increases. Table \ref{tab:3} shows the aggregate measures, $DI$ and $DC$, as well as the average connectivity, the
Gini coefficient for links concentration ($G$), Gini coefficient for distribution of credit links ($G_{in}$) and Gini coefficient for distribution
of debt links.

 \begin{table} [htbp]
     \caption[$DI$ e $DC$, connectivity and concentration for network $GD$, $S$ and $GC$ (types 0 and 1)]{Aggregate measures, $DI$ and $DC$, average connectivity,
     $<k>$, Gini coefficient for links distribution, $G$, Gini coefficient for distribution of credit links, $G_{in}$, and Gini coefficient for distribution
     of debt links, $G_{out}$, for network types 0 and 1.}
    \label{tab:3}
    \centering
    \begin{tabular}{l c c c c c c}
    \toprule
      & $DI$ & $DC$ & $<k>$ & $G$ & $G_{in}$ & $G_{out}$\\
    \toprule
      {\bf $GC_0$} & 0.433 ($\pm$0.001) & 0.494 ($\pm$0.023) & 2.652 ($\pm$0.028) & 0.456 ($\pm$0.008) & 0.748 ($\pm$0.011) & 0.410 ($\pm$0.008)\\[1.5mm]
      {\bf $S_0$} & 0.455 ($\pm$0.004) & 0.792 ($\pm$0.024) & 2.663 ($\pm$0.041) & 0.429 ($\pm$0.006) & 0.578 ($\pm$0.015) & 0.576 ($\pm$0.012)\\[1.5mm]
      {\bf $GD_0$} &  0.480 ($\pm$0.004) & 0.988 ($\pm$0.015) & 2.646 ($\pm$0.039) & 0.457 ($\pm$0.006) & 0.418 ($\pm$0.013) & 0.746 ($\pm$0.009)\\
      & & & & & & \\
      {\bf $GC_1$} & 0.426 ($\pm$0.001) & 0.331 ($\pm$0.028) & 7.406 ($\pm$0.166) & 0.630 ($\pm$0.008) & 0.805 ($\pm$0.010) & 0.561 ($\pm$0.010)\\[1.5mm]
      {\bf $S_1$} & 0.429 ($\pm$0.001) & 0.735 ($\pm$0.025) & 7.484 ($\pm$0.227) & 0.612 ($\pm$0.007) & 0.669 ($\pm$0.009) & 0.671 ($\pm$0.008)\\[1.5mm]
      {\bf $GD_1$} & 0.437 ($\pm$0.002) & 1.113 ($\pm$0.027) & 7.425 ($\pm$0.165) & 0.630 ($\pm$0.008) & 0.560 ($\pm$0.010) & 0.805 ($\pm$0.007)\\
    \toprule
    \end{tabular}
\end{table}

As we see, the increase in average connectivity when accompanied by an increase in the concentration of links have different effects depending on the
symmetry of the network. As we move from network $GC_0$ to $GC_1$ we see a decrease of contagion effect in response to the increased number of connections
and higher concentration of credits (concentration of in-links) that comes with the change in concentration. The average connectivity increase from 2.65
links per node to 7.41 and the Gini coefficient, which is 0.46 in $GC_0$, becomes 0.63. The same behavior we see in Gini coefficient of the distribution
of credit links (0.75 to 0.81) and in Gini of the distribution of debt links (0.41 to 0.56). The aggregate {\it Default Impact} does not suffer large
variation, varying from 0.43 in $GC_0$ to 0.42 in $GC_1$. On the other hand, the aggregate $DC$ ({\it Default Cascade}) has a significant decrease from
0.49 to 0.33.

We conclude that for networks $GC$, in which the credit concentration is larger than debt concentration, an increase in connectivity accompanied by an
increase in concentration of links has the effect of increasing resistance to contagion. Increasing credit concentration in nodes that become major creditors
of the system is certainly the factor responsible for the improvement in network resilience to contagion. Very connected creditors have many in-links, so
that its failure propagates less, only through its few out-links. Moreover, when a neighbor bank of the big creditor defaults, the transmitted loss
represents a small fraction of creditor bank total exposures, since it has many other counterparties. Here we see the positive results of having a major lender
that has diversified its risk among many counterparties.

For symmetric networks there is only a small improvement in resilience when we go from $S_0$ to $S_1$. As shown in table \ref{tab:3}, the
values of $DI$ and $DC$ undergo a slight reduction ($DI$ varying from 0.46 to 0.43 and $DC$ varying from 0.79 to 0.73).

For networks $GD$ the situation is ambiguous: increased connectivity accompanied by increased links concentration causes a slight reduction of $DI$, from
0.48 to 0.44. At the same time we see an increase in {\it Cascade Default}, with $DC$ varying from 0.99 to 1.11. For networks $GD$, the increase in
links concentration favors the appearance of large debtor banks. These banks have great potential for contagion, since most of their links are directed to
the system, having a destabilizing role in the network. The fact that $DC$ has increased with the
increase of connectivity and links concentration is a result of that effect. Even so, the concentration of links also increases the concentration of credit links
in creditor banks, which is a stabilizing factor that, with the increase in connectivity, causes decrease in $DI$.

\subsubsection{Increased connectivity and links concentration similar to type 0}

In type 2 we consider networks where the average connectivity is raised by increasing $\beta$ (from 0.25 to 0.75), while we raise $\delta_{in}$ and
$\delta_{out}$ in order to offset the trend of increasing concentration. We keep the 1:3 ratio between $\alpha$ and $\gamma$.

As we have seen in section \ref{cap3}, the parameters $\delta_{in}$ and $\delta_{out}$ represent probabilities distributed equally between nodes, giving
every node a chance of being selected in the attachment process. The preferential attachment concentrates links in large connected nodes, while the
parameters $\delta_{in}$ and $\delta_{out}$ can limit this tendency.

With $\beta$ increased to 0.75, we raise $\delta_{in}$ and $\delta_{out}$ 25 times
in order to maintain the Gini coefficient of the links distribution the closest to the value it has in type 0, i.e. around 0.45. The best approach we get
with $\beta=0,75$ is a Gini of 0.47. The parameter values used in types 0 and 2 are shown again in table \ref{tab:4} to facilitate comparison. Table
\ref{tab:5} shows the aggregate measures $DI$ and $DC$, and the values of average connectivity, concentration of links and partial concentrations (credit
links and debt links).

 \begin{table}[htbp]
     \caption[Parameters for the networks $GD$, $S$ e $GC$ (types 0 and 2)]{Parameter values ??used to generate the networks of types 0 and 2.}
    \label{tab:4}
    \vspace{0.5em}
    \centering
    \begin{tabular}{l c c c c c}
    \toprule
      & $\alpha$ & $\beta$ & $\gamma$ & $\delta_{in}$ & $\delta_{out}$\\
    \toprule
      {\bf $GC_0$} & 0.5625 & 0.2500 & 0.1875 & 1.00 & 3.00\\
      {\bf $S_0$} & 0.3750 & 0.2500 & 0.3750 & 2.00 & 2.00\\
      {\bf $GD_0$} & 0.1875 & 0.2500 & 0.5625 & 3.00 & 1.00\\[2.0mm]
      {\bf $GC_2$} & 0.1875 & 0.7500 & 0.0625 & 25.00 & 75.00\\
      {\bf $S_2$} & 0.1250 & 0.7500 & 0.1250 & 50.00 & 50.00\\
      {\bf $GD_2$} & 0.0625 & 0.7500 & 0.1875 & 75.00 & 25.00\\
    \toprule
    \end{tabular}
\end{table}

 \begin{table} [htbp]
     \caption[$DI$ e $DC$, connectivity and concentration for the networks $GD$, $S$ and $GC$ (types 0 and 2)]{Aggregate measures, $DI$ and $DC$, average
      connectivity, $<k>$, Gini coefficient of the distribution of links, $G$, Gini coefficient of the distribution of credit links, $G_ {in}$, and
      Gini coefficient of the distribution of debt links, $G_{out}$, for types 0 and 2.}
    \label{tab:5}
    \vspace{0.5em}
    \centering
    \begin{tabular}{l c c c c c c}
    \toprule
      & $DI$ & $DC$ & $<k>$ & $G$ & $G_{in}$ & $G_{out}$\\
    \toprule
      \vspace{0.5em}
      {\bf $GC_0$} & 0.433 ($\pm$0.001) & 0.494 ($\pm$0.023) & 2.652 ($\pm$0.028) & 0.456 ($\pm$0.008) & 0.748 ($\pm$0.011) & 0.410 ($\pm$0.008)\\
      \vspace{0.5em}
      {\bf $S_0$} & 0.455 ($\pm$0.004) & 0.792 ($\pm$0.024) & 2.663 ($\pm$0.041) & 0.429 ($\pm$0.006) & 0.578 ($\pm$0.015) & 0.576 ($\pm$0.012)\\
      \vspace{0.5em}
      {\bf $GD_0$} & 0.480 ($\pm$0.004) & 0.988 ($\pm$0.015) & 2.646 ($\pm$0.039) & 0.457 ($\pm$0.006) & 0.418 ($\pm$0.013) & 0.746 ($\pm$0.009)\\
      & & & & & & \\
      \vspace{0.5em}
      {\bf $GC_2$} & 0.428 ($\pm$0.001) & 0.574 ($\pm$0.023) & 7.813 ($\pm$0.200) & 0.476 ($\pm$0.006) & 0.555 ($\pm$0.008) & 0.465 ($\pm$0.007)\\
      \vspace{0.5em}
      {\bf $S_2$} & 0.429 ($\pm$0.001) & 0.723 ($\pm$0.021) & 7.812 ($\pm$0.141) & 0.471 ($\pm$0.008) & 0.507 ($\pm$0.009) & 0.506 ($\pm$0.008)\\
      \vspace{0.5em}
      {\bf $GD_2$} & 0.430 ($\pm$0.001) & 0.884 ($\pm$0.020) & 7.817 ($\pm$0.139) & 0.475 ($\pm$0.007) & 0.463 ($\pm$0.008) & 0.554 ($\pm$0.007)\\
    \toprule
    \end{tabular}
\end{table}

When comparing networks $GC_0$ and $GC_2$ we notice the increase in {\it Default Cascade}, $DC$, from 0.49 to 0.57. The default impact index, $DI$,
remains stable at 0.43.

Despite our attempt to isolate the connectivity effect by fixing Gini closest to the original value (0.45), we can not attribute the raising of the index
$DC$ only to the increased connectivity, because while the concentration of links remain close to the original value, there is a variation of partial
concentrations, $G_{in}$ and $G_{out}$. Indeed, when we change $\delta_{in}$ and $\delta_{out}$, we observe the increase of concentration of debt links in
large debtors nodes ($G_{out}$ varying from 0.41 to 0.46) and a reduction in concentration of credits ($G_{in}$ from 0.75 to 0.55). These changes contribute
to the worsening of index $DC$.

In the case of symmetric networks we see a slight improvement in indices: when comparing network $S_0$ and $S_2$ one can observe the variation of $DI$
from 0.46 to 0.43 and of $DC$ from 0.79 to 0.72.

For the network $GD$, an increases of connectivity slightly improves measure $DI$, from 0.48 to 0.43, and also improves the index $DC$, from 0.99 to 0.88.
Here also stands out, besides the increased connectivity, the variation of the partial concentrations, towards a decreased concentration of debts and
an increased concentration of credits.

\subsubsection{Increased connectivity and lower links concentration}

The type 3 repeats the experiment of Cont and Moussa \cite{Rama2010b} to test an increasing in connectivity. Type 3 network is built with same parameter
values as type 0. Subsequently its connectivity is increased by adding new links randomly distributed among its nodes. In this way, we have a more
connected and less concentrated network: the Gini coefficient goes from 0.45 to 0.24 and the partial Gini coefficients also suffer reduction. The data are
presented in table \ref{tab:6}.

 \begin{table} [htbp]
     \caption[$DI$ e $DC$, connectivity and concentration for the networks $GD$, $S$ and $GC$ (types 0 and 3)]{Aggregate measures, $DI$ and $DC$, average
      connectivity, $<k>$, Gini coefficient of the distribution of links, $G$, Gini coefficient of the distribution of credit links, $G_ {in}$, and
      Gini coefficient of the distribution of debt links, $G_{out}$, for types 0 and 3.}
    \label{tab:6}
    \vspace{0.5em}
    \centering
    \begin{tabular}{l c c c c c c}
    \toprule
      & $DI$ & $DC$ & $<k>$ & $G$ & $G_{in}$ & $G_{out}$\\
    \toprule
      \vspace{0.5em}
      {\bf $GC_0$} & 0.433 ($\pm$0.001) & 0.494 ($\pm$0.023) & 2.652 ($\pm$0.028) & 0.456 ($\pm$0.008) & 0.748 ($\pm$0.011) & 0.410 ($\pm$0.008)\\
      \vspace{0.5em}
      {\bf $S_0$} & 0.455 ($\pm$0.004) & 0.792 ($\pm$0.024) & 2.663 ($\pm$0.041) & 0.429 ($\pm$0.006) & 0.578 ($\pm$0.015) & 0.576 ($\pm$0.012)\\
      \vspace{0.5em}
      {\bf $GD_0$} &  0.480 ($\pm$0.004) & 0.988 ($\pm$0.015) & 2.646 ($\pm$0.039) & 0.457 ($\pm$0.006) & 0.418 ($\pm$0.013) & 0.746 ($\pm$0.009)\\
      & & & & & & \\
      \vspace{0.5em}
      {\bf $GC_3$} & 0.429 ($\pm$0.001) & 0.678 ($\pm$0.020) & 7.789 ($\pm$0.040) & 0.245 ($\pm$0.006) & 0.378 ($\pm$0.010) & 0.276 ($\pm$0.005)\\
      \vspace{0.5em}
      {\bf $S_3$} & 0.434 ($\pm$0.001) & 0.860 ($\pm$0.027) & 7.800 ($\pm$0.041) & 0.233 ($\pm$0.004) & 0.318 ($\pm$0.007) & 0.317 ($\pm$0.005)\\
      \vspace{0.5em}
      {\bf $GD_3$} & 0.450 ($\pm$0.005) & 1.007 ($\pm$0.017) & 7.794 ($\pm$0.043) & 0.256 ($\pm$0.005) & 0.277 ($\pm$0.005) & 0.374 ($\pm$0.007)\\
    \toprule
    \end{tabular}
\end{table}

As in types 1 and 2, all three networks ($GC$, $S$ e $GD$) show a slight improvement of aggregate impact measure, $DI$, suggesting that the
{\it Default Impact} decreases with increasing connectivity regardless if it is accompanied by an increase or decrease in links concentration. Nevertheless
the {\it Default Cascade} certainly depends on concentrations, as can be seen from the data presented. With increased connectivity and reduction of
concentration of links (including the partial concentrations), the network $GC$ suffers an increase of {\it Default Cascade}, with the aggregate measure
$DC$ varying from 0.49 to 0.68. The symmetric network also suffers deterioration in index $DC$, from 0.79 to 0.85. In turn the network $GD$ has the least
change in the index, from 0.99 to 1.01. Again the data suggest that changes in links concentrations are important to determine the {\it Default Cascade}.

\subsubsection{Same connectivity and lower links concentration}

Finally we tested the networks for a decrease in concentration of links, maintaining the same connectivity. We build networks of type 4 by
maintaining the same value of $\beta$ as the original networks, ie, beta = 0.25, and increasing by 10 times the values of $\delta_{in}$ and $\delta_{out}$.
Parameter values ??are presented in table \ref{tab:7}.

\begin{table}[htbp]
     \caption[Parameters for the networks $GD$, $S$ e $GC$ (types 0 and 4)]{Parameter values ??used to build the networks of types 0 and 4.}
    \label{tab:7}
    \vspace{0.5em}
    \centering
    \begin{tabular}{l c c c c c}
    \toprule
      & $\alpha$ & $\beta$ & $\gamma$ & $\delta_{in}$ & $\delta_{out}$\\
    \toprule
      {\bf $GC_0$} & 0.5625 & 0.2500 & 0.1875 & 1.00 & 3.00\\
      {\bf $S_0$} & 0.3750 & 0.2500 & 0.3750 & 2.00 & 2.00\\
      {\bf $GD_0$} & 0.1875 & 0.2500 & 0.5625 & 3.00 & 1.00\\[2.0mm]
      {\bf $GC_4$} & 0.5625 & 0.2500 & 0.1875 & 10.00 & 30.00\\
      {\bf $S_4$} & 0.3750 & 0.2500 & 0.3750 & 20.00 & 20.00\\
      {\bf $GD_4$} & 0.1875 & 0.2500 & 0.5625 & 30.00 & 10.00\\
    \toprule
    \end{tabular}
\end{table}

The indices of impact, as well as data connectivity and concentration are shown in table \ref{tab:8}.

 \begin{table} [htbp]
     \caption[$DI$ e $DC$, connectivity and concentration for the networks $GD$, $S$ and $GC$ (types 0 and 4)]{Aggregate measures, $DI$ and $DC$, average
      connectivity, $<k>$, Gini coefficient of the distribution of links, $G$, Gini coefficient of the distribution of credit links, $G_ {in}$, and
      Gini coefficient of the distribution of debt links, $G_{out}$, for types 0 and 4.}
    \label{tab:8}
    \vspace{0.5em}
    \centering
    \begin{tabular}{l c c c c c c}
    \toprule
      & $DI$ & $DC$ & $<k>$ & $G$ & $G_{in}$ & $G_{out}$\\
    \toprule
      \vspace{0.5em}
      {\bf $GC_0$} & 0.433 ($\pm$0.001) & 0.494 ($\pm$0.023) & 2.652 ($\pm$0.028) & 0.456 ($\pm$0.008) & 0.748 ($\pm$0.011) & 0.410 ($\pm$0.008)\\
      \vspace{0.5em}
      {\bf $S_0$} & 0.455 ($\pm$0.004) & 0.792 ($\pm$0.024) & 2.663 ($\pm$0.041) & 0.429 ($\pm$0.006) & 0.578 ($\pm$0.015) & 0.576 ($\pm$0.012)\\
      \vspace{0.5em}
      {\bf $GD_0$} &  0.480 ($\pm$0.004) & 0.988 ($\pm$0.015) & 2.646 ($\pm$0.039) & 0.457 ($\pm$0.006) & 0.418 ($\pm$0.013) & 0.746 ($\pm$0.009)\\
      & & & & & & \\
      \vspace{0.5em}
      {\bf $GC_4$} & 0.446 ($\pm$0.001) & 0.714 ($\pm$0.016) & 2.644 ($\pm$0.048) & 0.394 ($\pm$0.006) & 0.608 ($\pm$0.009) & 0.385 ($\pm$0.011)\\
      \vspace{0.5em}
      {\bf $S_4$} & 0.459 ($\pm$0.001) & 0.871 ($\pm$0.022) & 2.635 ($\pm$0.046) & 0.388 ($\pm$0.007) & 0.509 ($\pm$0.009) & 0.509 ($\pm$0.011)\\
      \vspace{0.5em}
      {\bf $GD_4$} & 0.466 ($\pm$0.002) & 1.016 ($\pm$0.018) & 2.661 ($\pm$0.036) & 0.394 ($\pm$0.007) & 0.383 ($\pm$0.013) & 0.607 ($\pm$0.010)\\
    \toprule
    \end{tabular}
\end{table}

For networks $GC$ and $S$ the lowest concentration of links causes worsening of the indices of contagion. For such networks the negative effect of the
reduction in concentration of credits is superior to the positive effect of reducing the concentration of debts.

For networks $GD$ the $DI$ index shows a slight improvement (from 0.48 to 0.46), whereas the index $DC$ has a slight worsening (0.99 to 1.01), indicating
that in this case the two effects are balanced.

\begin{figure}[htba]
    \centering
    \includegraphics[width=10cm]{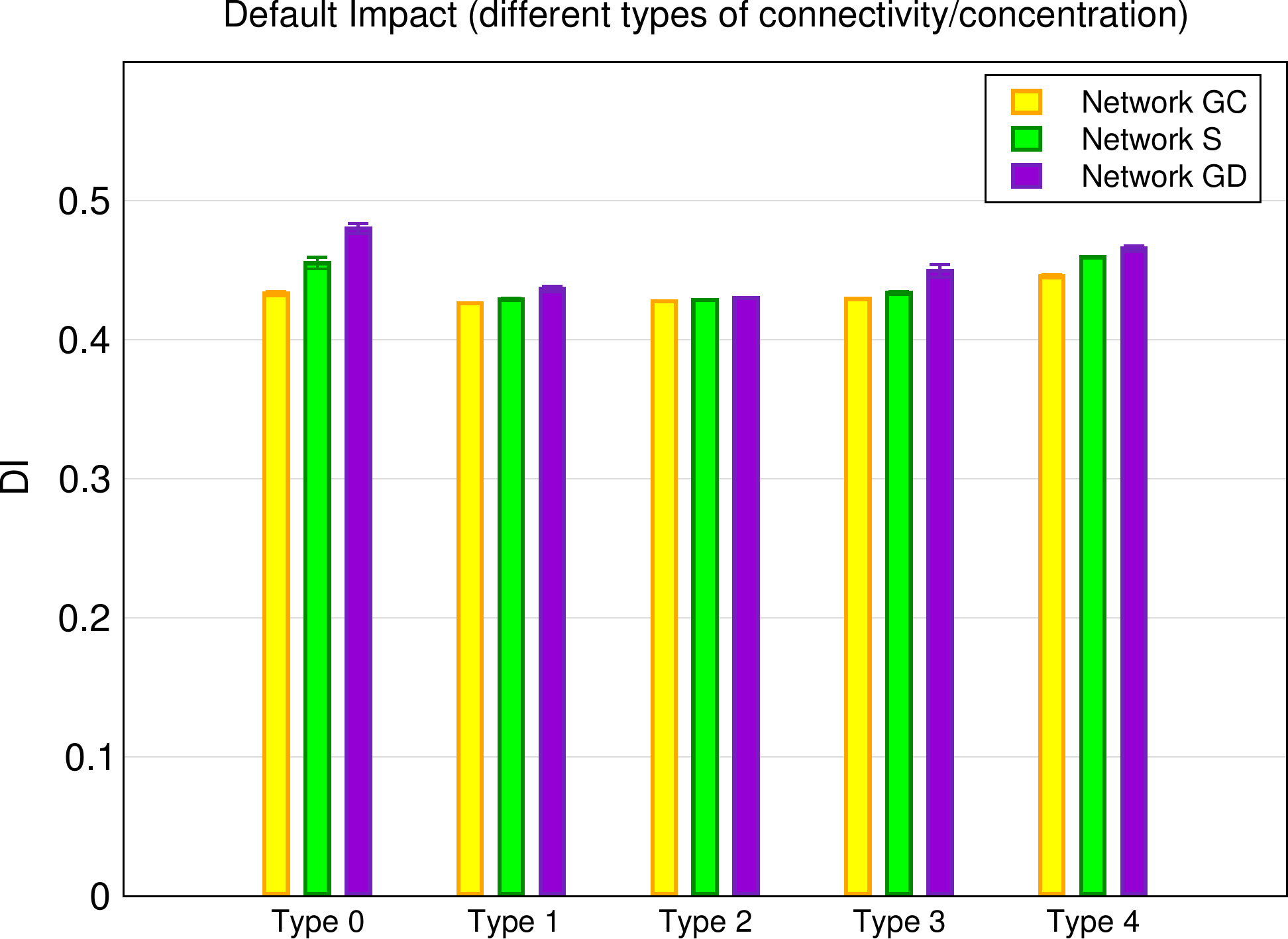}
    \caption[{\it Default Impact} (different types of connectivity/concentration)]{{\it Default Impact} (aggregate measure) for the 5 types of network
    analyzed.}
    \label{fig:Perfis_di}
\end{figure}

\begin{figure}[htba]
    \centering
    \includegraphics[width=10cm]{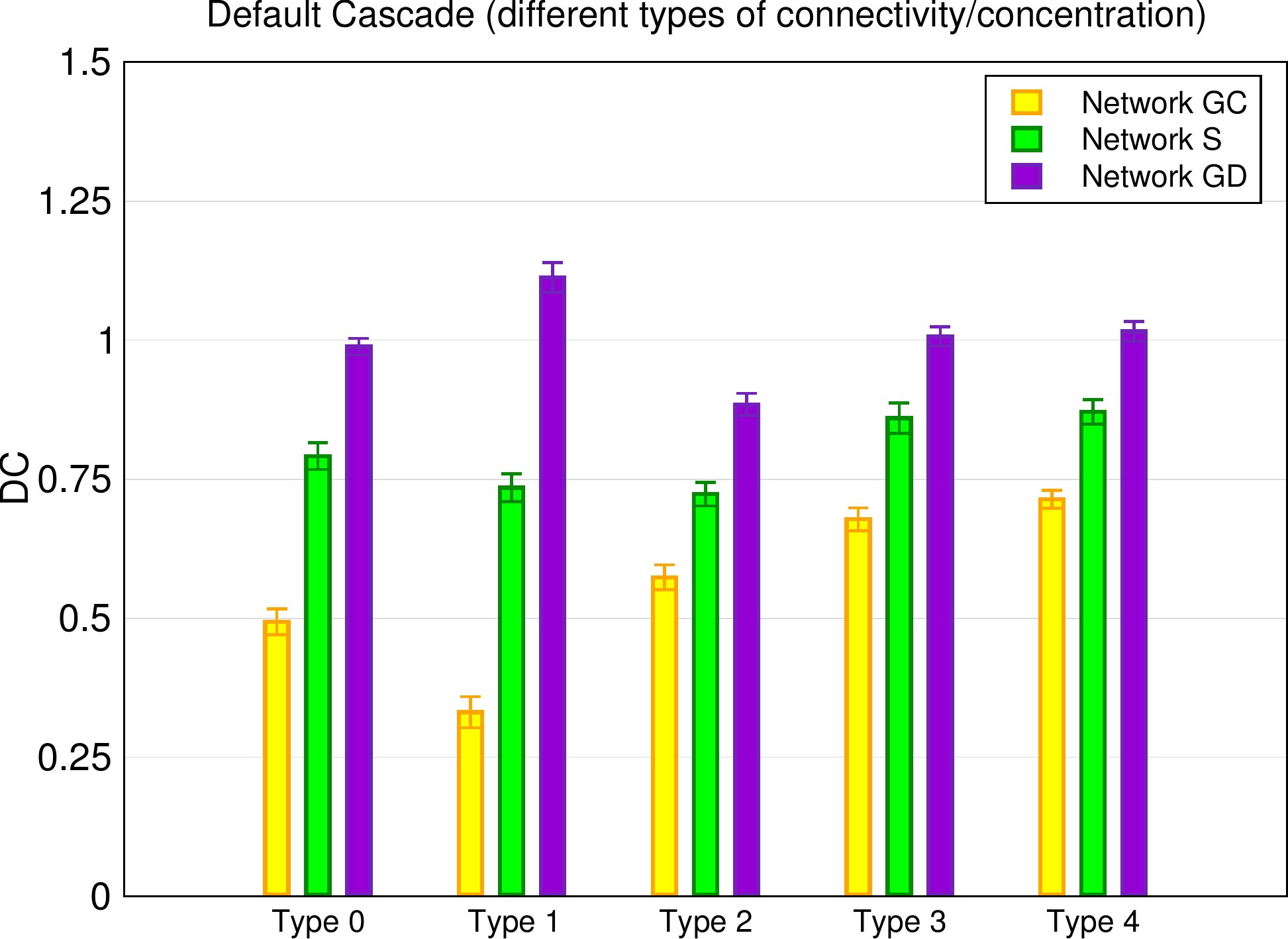}
    \caption[{\it Default Cascade} (different types of connectivity/concentration)]{{\it Default Cascade} (aggregate measure) for the 5 types of network
    analyzed.}
    \label{fig:Perfis_dc}
\end{figure}

Figures \ref{fig:Perfis_di} e \ref{fig:Perfis_dc} summarize the results of this section presenting the values of $DI$ e $DC$ for the 5 types of network
analyzed.

Comparisons among types suggests that for networks constructed with the algorithm of Bollobás et al. \cite{conf/soda/BollobasBCR03} and with exposures
positively related to connectivity, the best scenario is that of a more connected network with high concentration of credits, featuring large creditors
nodes which act as stabilizers of the network. It should be emphasized again that in comparisons among network types performed in this work, we do not
consider differences among nodes regarding their probabilities of default, differences that can change the evaluation of each network type.

\section{Conclusions}\label{cap7}

In this paper, we have analyzed the financial contagion via mutual exposures in the interbank market through simulations of networks whose degree
distributions follows power laws.

We have seen that among the measures of systemic importance ($DI$ and $DC$), {\it Default Cascade} ($DC$) is the one that most differentiates the categories
of network analyzed. We also observe that, for all categories, both the {\it Default Impact} and {\it Default Cascade} of each node alone does not reach
large percentage of the network assets and number of nodes, respectively. This result is in agreement with the results of stress tests on empirical
networks (Upper, C. \cite{Upper2011}).
Although contagion generated by the failure of an individual node has been shown to be of small amplitude, the differences found among network types may
be relevant in the event of market shocks that affect capital of several institutions simultaneously, which would make the system more vulnerable to
contagion.

Comparisons among types of network suggest that, for networks whose distributions are close to power laws and exposure is positively related to
connectivity, the best scenario is one with a more connected network with high concentration of credits, featuring large creditors nodes which act as
stabilizers of the network. This results suggests that the asymmetry observed in distributions of certain real networks is a positive factor, as long as
the network be more concentrated in distribution of credits ({\it in} links). In real networks having power laws exponents
estimated between 2 and 3, the results imply that the most stable networks are those with distribution of {\it in} links with exponent 2 and
distribution of {\it out} links with exponent 3.

As expected, the increase of capital level reduces contagion in all three categories of network, since the equity of banks absorbs impacts
suffered by them. We also observed that larger networks have nodes with less potential for contagion, situation already observed in
previous works (Cont and Moussa \cite{Rama2010b}).

Our simulation shows that the {\it default impact} is strongly determined by the {\it initial impact}, which was expected, since the assets of large banks
(greater initial impact) represent significant portion of the network assets. However, the {\it initial impact} does not determines the
{\it default cascade} to the same extent: {\it default cascade} appears to be more sensitive to characteristics of local connectivity.

The results suggest that the size of the balance sheet is the most important factor in determining the impact on assets resulting from the failure of a
node, and should not be disregarded or replaced by topological measures that reflect only information of network connectivity. On the other hand, the
network structure has important consequences on the {\it default cascade}. In some cases, the banks which trigger the largest cascades
are not the ones with greater balance sheet. For those banks, topological measures such as {\it Local network frailty} are good indicators of their
systemic importance.



\end{document}